\documentclass[twocolumn,twocolappendix,tighten,times,astrosymb]{aastex631}

\usepackage{mathtools}

\usepackage[T1]{fontenc}
\usepackage{microtype}

\shorttitle{\footnotesize Gro\v selj, Sironi, \& Spitkovsky}
\shortauthors{\footnotesize Gro\v selj, Sironi, \& Spitkovsky}
\begin{document}

\title{\large Long-term Evolution of Relativistic Unmagnetized Collisionless Shocks}

\author[0000-0002-5408-3046]{Daniel Gro\v selj}
\affiliation{Centre for mathematical Plasma Astrophysics, Department of Mathematics, 
KU Leuven, B-3001 Leuven, Belgium}
\affiliation{Department of Astronomy and Columbia Astrophysics Laboratory, 
Columbia University, New York, NY 10027, USA}
\correspondingauthor{Daniel Gro\v selj}
\email{daniel.groselj@kuleuven.be}

\author[0000-0002-1227-2754]{Lorenzo Sironi}
\affiliation{Department of Astronomy and Columbia Astrophysics 
Laboratory, Columbia University, New York, NY 10027, USA}
\affiliation{Center for Computational Astrophysics, Flatiron Institute, 162 5th Avenue, New York, NY 10010, USA}

\author[0000-0001-9179-9054]{Anatoly~Spitkovsky}
\affiliation{Department of Astrophysical Sciences, Princeton University, 
Princeton, NJ 08544, USA}

\begin{abstract}
We study a relativistic collisionless electron-positron shock propagating
into an unmagnetized ambient medium using 2D particle-in-cell 
simulations of unprecedented duration and size. The shock generates 
intermittent magnetic structures of increasingly larger size as the simulation progresses. 
Toward the end of our simulation, at around 26,000 plasma times, 
the magnetic coherence scale approaches $\lambda\sim 100$ plasma skin depths, both 
ahead and behind the shock front. We anticipate a continued growth of $\lambda$ beyond 
the time span of our simulation, as long as the shock accelerates particles to increasingly higher energies.
The post-shock field is concentrated in localized patches, which maintain a 
local magnetic energy fraction $\varepsilon_B\sim 0.1$. 
Particles randomly sampling the downstream fields 
spend most of their time in low field regions ($\varepsilon_B\ll 0.1$),
but emit a large fraction of the synchrotron power in the localized patches
with strong fields ($\varepsilon_B\sim 0.1$). 
Our results have important implications for models of gamma-ray burst afterglows.
\end{abstract}


\section{Introduction}
Gamma-ray bursts (GRBs) are powerful cosmic explosions driven by the collapse of a massive star or 
the merger of binary neutron stars \citep{Meszaros2002,Piran2004,KumarP2015}. The prompt emission 
stemming from the internal dissipation of the GRB jet is followed by an afterglow powered by 
the external relativistic collisionless shock, which expands into the surrounding medium. The afterglow
unfolds from a few seconds up to decades after the explosion, as we are now witnessing for 
GRB 221009A, the ``brightest of all time'' \citep{Burns2023,Williams2023,Laskar2023}. The emission
from GRB afterglows is distinctly nonthermal and broadband. It ranges from the radio to the gamma-ray band
\citep{Abdo2009, Ackermann2010, DePasquale2010, Chandra2012, Ackermann2014, Perley2014}, and has been recently observed 
even at TeV energies 
\citep{HESS2019,HESS2021,MAGIC2019,MAGIC2019b, Huang2022TeV,LHAASO2023}.

\begin{figure*}[htb!]
\includegraphics[width=\textwidth]{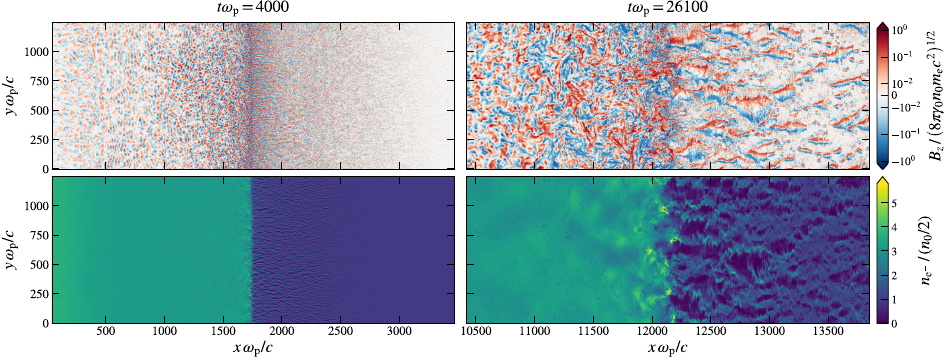}
\caption{\label{fig:structure} Structure of a collisionless relativistic pair plasma shock, propagating into an 
unmagnetized upstream medium with a bulk Lorentz factor $\gamma_0=10$, at $t\omega_{\rm p} = 4000$ (left) and at $t\omega_{\rm p} = 26100$ (right). 
On the top we show the out-of-plane magnetic field $B_z$, and on the bottom the electron number density $n_{\rm e^{-}}$. An animated version of this figure is available online. The animation
lasts 40 s and shows the spatiotemporal 
evolution of $B_z$ (top panel) and $n_{\rm e^{-}}$ (bottom panel) from the start ($t\omega_{\rm p}=0$)
to the end ($t\omega_{\rm p} = 26100$) of the simulation.}
\end{figure*}

GRB afterglow shocks propagate into a very weakly magnetized ambient medium. The ratio of the 
magnetic to the particle energy density of the cold ambient medium (i.e., the magnetization) is $\sigma\lesssim 10^{-5}$ 
for massive stellar progenitors \citep{Crowther2007}, and could be as low as $\sigma\sim 10^{-9}$ if the shock
expands into an interstellar-like medium \citep[e.g.,][]{Sironi2013,Groselj2022}.
A mechanism for magnetic field generation must be present at the shock in order to facilitate 
efficient particle scattering and their acceleration via the Fermi process
\citep[e.g.,][]{Blandford1978,Bell1978,Drury1983,Blandford1987, Achterberg2001}, and to explain the resulting broadband emission.
Kinetic particle-in-cell (PIC) 
simulations of relativistic collisionless shocks 
\citep{Spitkovsky2008, Spitkovsky2008b, Martins2009, Nishikawa2009, Haugbolle2011, Sironi2013}
have shown that the Weibel filamentation instability \citep{Fried1959,Weibel1959, Medvedev1999, Silva2003, 
Achterberg2007b, Bret2014, Takamoto2018, Lemoine2019} generates the required fields, 
which serve as the scattering agents for the Fermi process. 
However, Weibel-generated fields grow on plasma microscales and decay quickly 
downstream from the shock \citep{Gruzinov2001, Chang2008, Lemoine2015}, which 
is at odds with the relatively high field strengths inferred for some 
GRB afterglows \citep[e.g.,][]{Wijers1999,Panaitescu2002}, including the recent GRB 221009A \citep{Laskar2023}.
Moreover, particle scattering in microscale fields presents a challenge for the 
acceleration to very high energies \citep{Kirk2010,Sironi2013, Reville2014, Asano2020, Huang2022}.
The largest PIC simulations reported in the literature \citep{Keshet2009} found that the acceleration of 
particles to higher energies drives the formation of magnetic fields on progressively larger scales over time, 
which slows down the decay of the post-shock magnetic fields.
However, despite of the computational progress, previous as well as present PIC simulations 
need to be extrapolated to orders of magnitude longer time scales to make direct contact with observations, 
and so the long-term evolution of the shock remains a subject worth investigating.

In this Letter, we revisit the microphysics of relativistic unmagnetized
collisionless shocks using PIC simulations of unprecedented duration and size. We observe the  
generation of intense magnetic structures in the upstream flow, reaching sizes up to $\sim 100$ 
plasma skin depths. The bulk of the post-shock plasma becomes magnetized, 
which leads to particle trapping in the magnetostatic turbulence, and prevents the fast decay 
of magnetic fields predicted for untrapped particles \citep{Chang2008,Lemoine2015}.
The downstream 
field is concentrated in localized patches, which occupy roughly one percent of the total 
volume but contain about half of the magnetic energy. Our results have direct implications 
for particle acceleration and emission from GRB afterglow shocks.

\section{Method}
We perform 2D simulations of relativistic collisionless shocks using the PIC code 
\textsc{osiris 4.0} \citep{Fonseca2002,Fonseca2013}. In order to evolve the simulations for as long as possible 
we employ for simplicity an electron-positron pair composition.
As is routinely done \citep[e.g.,][]{Sironi2013}, the shock formation is driven by the
reflection of the cold upstream flow (moving to the left) from 
the left simulation boundary at $x=0$. The
rest frame of the simulation coincides with the downstream frame and the shock propagates
in the positive $x$ direction. In order to reduce reflecting boundary artifacts, 
the initial upstream flow Lorentz factor transitions linearly from unity to the far-upstream value $\gamma_0=10$ 
over a distance of 400 $c/\omega_{\rm p}$ from $x=0$, 
where $c/\omega_{\rm p} = (\gamma_0 m_{\rm e}c^2 / 4\pi n_0 e^2)$ 
is the plasma skin depth and $n_0$ is the simulation-frame density of the upstream electrons and positrons. At $t=0$, the upstream
plasma fills the transition layer between $x=0$  and $x = 400\,c/\omega_{\rm p}$. As the simulation progresses, 
the region $x > 400\,c/\omega_{\rm p}$ is gradually populated with upstream plasma introduced from a moving particle injector, which 
moves to the right at light speed $c$ until it reaches the end of the box.

Our fiducial simulation has a box size of 
$20000 \times 1248\,c/\omega_{\rm p}$ and is evolved up to $t \omega_{\rm p}= 26100$. The cell size is $\Delta x = 0.2\,c/\omega_{\rm p}$ 
and our time step $\Delta t = 0.5\Delta x/c$. The upstream plasma is represented with 144 particles per cell per species 
with cubic spline shapes. The total number of particles at the end of the simulation exceeds 0.4 trillion.
Particle noise is further reduced by low-pass filtering the electric currents at
each step. The fields are advanced using 
a modified stencil \citep{Blinne2018} with improved numerical dispersion of electromagnetic 
waves to suppress the numerical Cherenkov instability \citep[see][]{Groselj2022}. 
For improved computational performance, we employ a vectorized particle 
push and deposit, together with a hybrid MPI/OpenMP 
parallelization strategy, all of which are readily available in the \textsc{osiris} code \citep{Fonseca2013}.

\section{Shock evolution}
\label{sec:evolution}

Figure~\ref{fig:structure} shows the evolution of the shock structure in our
fiducial simulation, as observed through the snapshots of the out-of-plane magnetic field (top panels)
and the electron density (bottom). The particles returning from the shock into the upstream flow 
drive plasma streaming instabilities, which grow and amplify magnetic fields. The upstream 
region populated with the self-generated fields defines a turbulent precursor, which expands with time as the returning particles 
propagate further into the upstream.
At early times ($t\omega_{\rm p}\lesssim 4000$; left panels in Fig.~\ref{fig:structure}), the upstream plasma near the shock front
develops a microscale filamentary structure, typical of shocks 
mediated by the Weibel instability \citep[e.g.,][]{Lemoine2019}. In contrast, at later stages 
of the shock evolution (right panels in Fig.~\ref{fig:structure}), 
the upstream Weibel filaments undergo a secondary nonlinear instability 
\citep{Honda2000, Peterson2021,Peterson2021b,Groselj2022}, which results in large-scale
magnetized plasma cavities.\footnote{Magnetized plasma cavities can be also
spotted in earlier simulations of relativistic pair shocks 
(see Fig.~1(b) from \citet{Keshet2009}). However, their role in the long-term evolution of pair shocks
was not fully appreciated at the time.} 

\begin{figure}[htb!]
\centering
\includegraphics[width=\columnwidth]{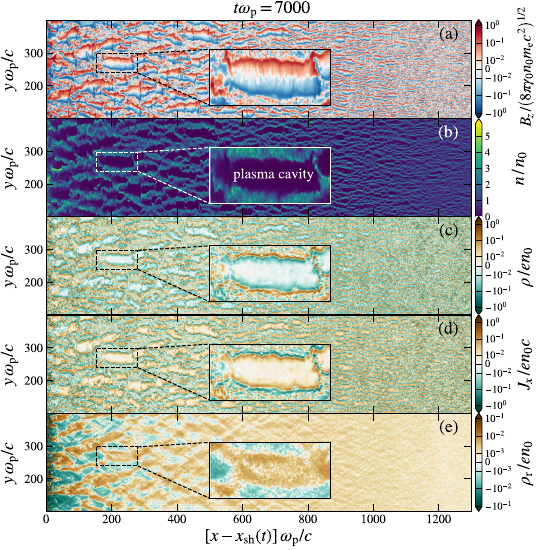}
\caption{Structure of the shock precursor populated with magnetized plasma cavities at $t\omega_{\rm p}=7000$. Only a fraction of the 
box is shown in the range $100<y\omega_{\rm p}/c<400$ and $0 < [x\! -\! x_{\rm sh}(t)]\omega_{\rm p}/c < 1300$, 
where $x_{\rm sh}(t)$ is the location of the propagating shock front.
From top to bottom we show the $z$ component of the magnetic field $B_z$ (a), the particle number density $n$ (b), 
the total electric charge density $\rho$ (c), the $x$ component of the electric current $J_x$ (d), 
and the charge density $\rho_{\rm r}$ of the returning particles with $\gamma\beta_x > 0$ (e). The inset plots
zoom in on the structure of a single cavity.\label{fig:cavities} }
\end{figure}

The structure of the precursor in the plasma cavity-dominated phase is illustrated in Fig.~\ref{fig:cavities}. 
Cavity formation is triggered by the development of a net electric current and charge in the 
returning population of electrons and positrons (defined as having $\gamma\beta_x > 0$), which stream against the
incoming flow (Fig.~\hyperref[fig:cavities]{\ref*{fig:cavities}(e)}). For example, at the time of Fig.~\ref{fig:cavities}, the returning particles 
carry an excess positive charge.
The asymmetry of the returning particles is then imprinted onto the current filaments 
of positive ($J_x > 0$) or negative ($J_x < 0$) polarity in the incoming background plasma (Fig.~\hyperref[fig:cavities]{\ref*{fig:cavities}(d)}), which
tries to neutralize the returning-particle charge and current.
The current filaments with the same polarity as the returning-particle current
($J_x > 0$ in Fig.~\hyperref[fig:cavities]{\ref*{fig:cavities}(d)}) become unstable 
to the cavitation instability and expand in the transverse ($y$) direction; the filaments of the opposite polarity ($J_x < 0$ in Fig.~\hyperref[fig:cavities]{\ref*{fig:cavities}(d)}) do not expand.
In the rest frame of the simulation, the electric current inside the cavities is sustained by a 
low-density population of incoming particles of the opposite charge sign to the
returning-particle charge density (Figs.~\hyperref[fig:cavities]{\ref*{fig:cavities}(c)}--\hyperref[fig:cavities]{\ref*{fig:cavities}(e)}). The current inside the cavities is left essentially unscreened, 
which enables the generation of intense magnetic fields \citep{Groselj2022}.
The newly born plasma cavities then grow and merge as they propagate toward the shock (Fig.~\hyperref[fig:cavities]{\ref*{fig:cavities}(b)}),  
which allows the magnetic field to cascade to progressively larger scales (Fig.~\hyperref[fig:cavities]{\ref*{fig:cavities}(a)}). 
At the end of our simulation, the largest individual cavities reach transverse sizes of about 
200 plasma skin depths (Fig.~\ref{fig:structure}, right panels).

\begin{figure}[htb!]
\centering
\includegraphics[width=\columnwidth]{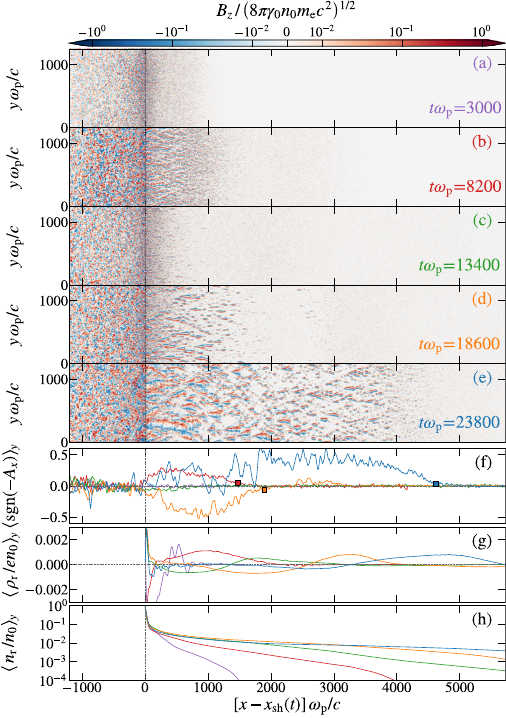}
\caption{\label{fig:evolution} Evolution of the magnetic field and the emergence of symmetry breaking.
Panels (a)--(e) show snapshots of $B_z$ at five times in the simulation. In panel (f) we show the
``order parameter'' $Q(x) = \langle{\rm sgn}(-A_x)\rangle_{\!y}$. Values of $Q(x)$ significantly above or below zero
indicate the presence of cavities of a given magnetic polarity (see text for details).
Panel (g) shows the transversely averaged upstream charge density of the returning particles with $\gamma\beta_x > 0$.
In (h) we show the transversely averaged number density of the returning particles.
Squares in (f) show the location of cavity birth (see Appendix~\ref{sec:symmetry}).}
\end{figure}

The current and charge of the returning particles, 
which drives cavity formation, exhibits oscillations in sign over time and in space. This gives rise to bursts (or cycles) of cavity generation with alternating polarity, as demonstrated
in Fig.~\ref{fig:evolution}. The returning-particle charge fluctuations originate near the shock with a time-varying sign and 
spread into the upstream (Fig.~\hyperref[fig:evolution]{\ref*{fig:evolution}(g)}). The amplitude of the charge 
fluctuations amounts to a significant fraction of $en_{\rm r}$, where $n_{\rm r}$ is the returning-particle 
number density (Fig.~\hyperref[fig:evolution]{\ref*{fig:evolution}(h)}).
In response to the charge oscillations in the stream of returning particles, the polarity of the magnetized plasma cavities 
changes over time (Figs.~\hyperref[fig:evolution]{\ref*{fig:evolution}(a)}--\hyperref[fig:evolution]{\ref*{fig:evolution}(e)}).
To quantify the symmetry breaking in the induced magnetic field structure we introduce an 
``order parameter'' $Q(x)$, defined as $Q(x) = \langle{\rm sgn}(-A_x)\rangle_{\!y}$, where
$A_x$ is the longitudinal component of the magnetic vector potential (Fig.~\hyperref[fig:evolution]{\ref*{fig:evolution}(f)}). 
We solve for $A_x$ in the Coulomb gauge, so that $-A_x$ can be related via Ampere's law 
to the current as $\nabla^2 (-A_x) = (4\pi /c) J_x$.
Essentially, $Q(x)$ measures the relative difference in the volume occupied by magnetic structures with $A_x < 0$ versus 
those with $A_x > 0$, and as such it provides a convenient diagnostic for tracking the changes in polarity over time and in space.
In our fiducial run, the first cavity cycle develops a positive polarity (see Fig.~\hyperref[fig:evolution]{\ref*{fig:evolution}(b)} and the corresponding profile of $Q(x)$). 
It starts around $t\omega_{\rm p}\approx 4500$ and lasts until $t\omega_{\rm p}\approx 11000$ or so.
The second cycle displays a negative polarity (Fig.~\hyperref[fig:evolution]{\ref*{fig:evolution}(d)}). It starts around $t\omega_{\rm p}\approx 13500$, 
continues up to $t\omega_{\rm p}\approx 23000$, and transitions into the third cycle, which features
cavities of positive polarity (Fig.~\hyperref[fig:evolution]{\ref*{fig:evolution}(e))}. The third cavity cycle is still ongoing 
at the end of the run (Fig.~\ref{fig:structure}, right panels).

The origin of the returning-particle charge and current oscillations is connected to the mechanism of
particle reflection at Weibel mediated shocks, which was recently studied in \citet{Parsons2023}. The authors found that 
returning particles of a given species (electrons or positrons) 
need to ``find'' current channels with a matching sign ($J_x<0$ for electrons 
and $J_x>0$ for positrons when the shock travels in the positive $x$ direction, as in our run) in order to efficiently 
move away from the shock into the upstream. 
This is supported by Figs.~\hyperref[fig:cavities]{\ref*{fig:cavities}(d)}--\hyperref[fig:cavities]{\ref*{fig:cavities}(e)}, 
which show that the local sign of the returning-particle charge density $\rho_{\rm r}$ correlates with the local sign 
of the current $J_x$, within a few hundred $c/\omega_{\rm p}$ from the shock.
The injection of particles back into the upstream through the current channels 
provides a mechanism for spontaneous symmetry breaking.
Namely, if the distribution of the background plasma current develops a seed asymmetry, 
 one of the returning species finds current channels with a matching sign 
more easily and is more successful at moving away from the shock, and so the returning population as a whole develops a net current and charge (Fig.~\hyperref[fig:cavities]{\ref*{fig:cavities}(e)}).
The returning particles of the species that sustains the excess charge then propagate further into the upstream and drive the formation
of cavities, which amplifies the original seed asymmetry of the background plasma current, reinforcing the positive feedback loop.
However, the continuous promotion of returning particles with a given charge sign generates 
a growing electrostatic potential between the shock and the far upstream. The 
electrostatic field pulls the species that provides the excess charge toward the 
shock and feeds back negatively on the asymmetry. Instead of continuous generation of cavities with a fixed polarity we 
therefore expect oscillations with a time-varying polarity, which are indeed
observed in our simulations. Note that when the composition includes ions, the difference in inertia between electrons and ions acts as a 
natural mechanism of symmetry breaking \citep{Lemoine2011,Groselj2022}. Thus, the exact nature of cavity cycles in electron-ion shocks could differ
from that in pair shocks.\footnote{We performed limited-duration electron-ion shock simulations and still found 
reversals of cavity polarity over time. However, the comparatively short duration of these simulations prevents us from making reliable extrapolations to longer times.}

\section{Magnetic field statistics}
\label{sec:stat}

\begin{figure}[htb!]
\centering
\includegraphics[width=\columnwidth]{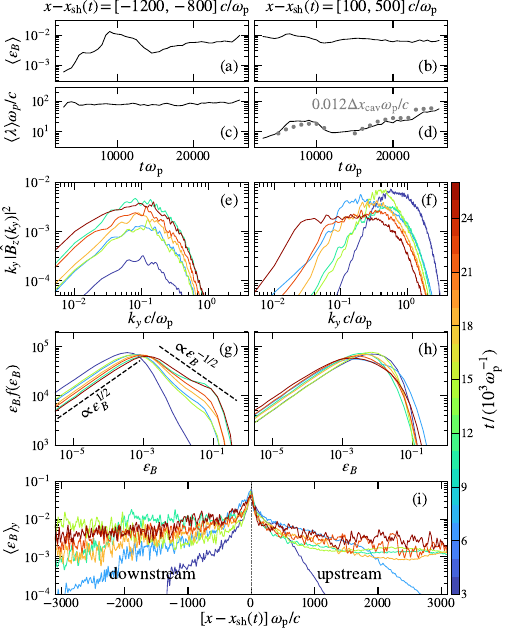}
\caption{\label{fig:statistics} Evolution of the magnetic field behind (left) and ahead of (right) the shock. 
Panels (a)--(d) show the average magnetic energy fraction 
$\varepsilon_B = B_z^2 / 8\pi \gamma_0n_0 m_{\rm e}c^2$ and the field transverse coherence scale $\lambda$
as a function of time in a fixed downstream ($x\!-\!x_{\rm sh} = [-1200, -800]\,c/\omega_{\rm p}$) and 
upstream ($x\!-\!x_{\rm sh} = [100,500]\,c/\omega_{\rm p}$) slab.
 Panels (e)--(f) show the magnetic spectrum as a 
function of the transverse wavenumber $k_y$ in the downstream (e) and 
upstream (f) slab (colors represent time). The spectrum is compensated by $k_y$ to 
highlight the energy content per scale.
In (g)--(h) we show the distribution 
of $\varepsilon_B$ in the downstream (g) and upstream (h) slab.
Finally, in (i) we show the transversely averaged profile of $\varepsilon_B$. Gray dots in panel (d) measure the distance $\Delta x_{\rm cav}\omega_{\rm p}/c$ 
(rescaled by a numeric prefactor $\approx 0.012$) between the shock and the upstream location of cavity birth (see Appendix~\ref{sec:symmetry} for details).}
\end{figure}

In Fig.~\ref{fig:statistics} we characterize the evolution of the magnetic 
field behind (left column) and ahead of (right column) the shock. Owing to cavity cycles (see Sec.~\ref{sec:evolution}), 
the evolution of magnetic field properties is not strictly monotonic. On the other hand, well-defined trends in the evolution can still be identified when the full time span
of the simulation is considered. In particular, the magnetic field transverse 
coherence scale\footnote{We define 
$\lambda = \pi\int k_y^{-1}P_B(k_y){\rm d}k_y / \int P_B(k_y){\rm d}k_y$,
where $k_y$ is the transverse wavenumber and $P_B(k_y)=|\hat B_z(k_y)|^2$ is the 1D magnetic energy
spectrum as a function of $k_y$.} $\lambda$ 
in front of the shock (Fig.~\hyperref[fig:statistics]{\ref*{fig:statistics}(d)}) grows by an order of magnitude over the duration of the simulation and 
approaches $\lambda\sim 100\, c/\omega_{\rm p}$ at late 
time. The corresponding late-time magnetic spectrum (Fig.~\hyperref[fig:statistics]{\ref*{fig:statistics}(f)})
is broadband; the magnetic energy uniformly fills the transverse wavenumber range $0.02 \lesssim k_y c/\omega_{\rm p} \lesssim 0.5$.
The size of structures in front of the shock correlates with the distance $\Delta x_{\rm cav}$ 
between the shock and the upstream location where the cavities arriving at the shock are born. 
In Fig.~\hyperref[fig:statistics]{\ref*{fig:statistics}(d)} we compare 
measurements of $\Delta x_{\rm cav}$ (gray dots) with the evolution of $\lambda$ and find empirically an approximately 
linear scaling $\lambda \approx 0.012 \Delta x_{\rm cav}$.\footnote{For details
on the $\Delta x_{\rm cav}$ measurement see Appendix~\ref{sec:symmetry}.} 
We discuss the possible implications of this result in Sec.~\ref{sec:discussion}.

Magnetic structures generated over the precursor are compressed at the shock and transmitted into the
downstream. Structures of growing size ahead of the shock thus imply 
that more energy is transferred to
longer wavelength modes in the downstream, which results in a slower overall 
decay of the magnetic energy behind the shock (Figs.~\hyperref[fig:statistics]{\ref*{fig:statistics}(a)} and ~\hyperref[fig:statistics]{\ref*{fig:statistics}(i)}).
The post-shock fields at wavenumbers $k_y \gtrsim \omega_{\rm p}/c$ 
decay within a few hundred skin depths from the shock, 
leaving behind the field at $k_y \lesssim 0.1\,\omega_{\rm p}/c$ 
(Fig.~\hyperref[fig:statistics]{\ref*{fig:statistics}(e)}), 
where the decay is slow because the particles are marginally magnetized. 
The Larmor radius $R_{\rm L}(\gamma)$ of a particle 
with typical Lorentz factor $\gamma\sim\gamma_0$
is $R_{\rm L}(\gamma_0) \simeq (2\varepsilon_B)^{-1/2}\,c /\omega_{\rm p}$, where 
$\varepsilon_B = B_z^2 / 8\pi \gamma_0n_0 m_{\rm e}c^2$ is the 
magnetic energy fraction. The mean of $\varepsilon_B$, measured in the same downstream slab as the magnetic spectrum, 
is $\langle\varepsilon_B\rangle\gtrsim 10^{-3}$ (Fig.~\hyperref[fig:statistics]{\ref*{fig:statistics}(a)}), 
except at early times ($t\omega_{\rm p}\lesssim 4000$). For $\langle\varepsilon_B\rangle\sim 10^{-3}$ particles
with $\gamma\sim\gamma_0$ become marginally magnetized at a 
wavenumber $k_y\sim \pi / R_{\rm L}(\gamma_0)  \sim 0.1\,\omega_{\rm p}/c$, which is near the peak of the magnetic spectrum 
(Fig.~\hyperref[fig:statistics]{\ref*{fig:statistics}(e)}). This confirms that the bulk of the post-shock plasma becomes magnetized in the long-time regime.
A similar finding was reported by \citet{Keshet2009} in a simulation evolved up to $t\omega_{\rm p}\approx 12600$.

The decaying magnetic field behind the shock evolves into a patchy configuration composed of 
long-lived structures with strong fields (reaching values up to $\varepsilon_B\sim 0.1$) 
and a background with weaker fields ($\varepsilon_B \ll 0.1$). This is reflected in the measured probability distribution
of $\varepsilon_B$ behind the 
shock (Fig.~\hyperref[fig:statistics]{\ref*{fig:statistics}(g)}). The 
distribution (with logarithmic bins for $\varepsilon_B$) 
rises as $\varepsilon_B f(\varepsilon_B) \propto \varepsilon_B^{1/2}$ on the low energy side toward the peak around 
$\varepsilon_B \sim 10^{-3}$. The $1/2$ slope is consistent with a normal Gaussian
distribution for the fluctuating field $B_z$.\footnote{If $f(b)\propto \exp(-a b^2)$, where 
$b = B_z / (8\pi\gamma_0 n_0 m_{\rm e}c^2)^{1/2}$, $b^2 = \varepsilon_B$ and $a$ is a constant,
then $\varepsilon_B f(\varepsilon_B) \propto \varepsilon_B^{1/2}\exp(-a\varepsilon_B)$.}
Due to intermittent magnetic structures, the distribution does not cut off exponentially above the
peak but instead transitions into a hard power-law tail with a slope around $-1/2$ (i.e., $\varepsilon_Bf(\varepsilon_B)\propto\varepsilon_B^{-1/2}$). The tail 
extends up to $\varepsilon_B\sim 0.1$ at late times.

\begin{figure}[htb!]
\centering
\includegraphics[width=\columnwidth]{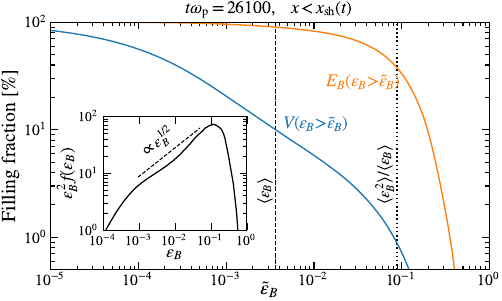}
\caption{\label{fig:sync} Fraction of the volume (blue) and of the magnetic energy (orange) contained in 
downstream ($x<x_{\rm sh}$) regions with
$\varepsilon_B$ larger than a given threshold $\tilde\varepsilon_B$. The value 
$\varepsilon_{B*} = \langle\varepsilon_B^2\rangle/\langle\varepsilon_B\rangle$ (dotted vertical line) 
is much greater than the mean $\langle\varepsilon_B\rangle$ (dashed line); $\varepsilon_{B*}$ represents the typical local 
energy density of intense magnetic fluctuations, which make a significant contribution to the synchrotron emission (see text for details). 
The inset shows the magnetic energy content (arbitrary units) 
per logarithmic interval in $\varepsilon_B$.}
\end{figure}

The relatively hard power-law tail above the peak of the probability distribution of $\varepsilon_B$ 
in the downstream has an immediate implication for models of GRB afterglows. Namely, a significant fraction of the 
synchrotron power $P_{\rm syn}\propto \gamma^2\varepsilon_B$ from particles randomly sampling the downstream field will be emitted at the localized structures 
with $\varepsilon_B\sim 0.1$. This is demonstrated in Fig.~\ref{fig:sync},
which shows the fraction of magnetic energy and the volume fraction contained in downstream regions with $\varepsilon_B$
larger than a given threshold $\tilde\varepsilon_B$.  Roughly $\sim 50$\% of the total magnetic energy, and 
hence of the total synchrotron power, is contributed by structures
with $\varepsilon_B \gtrsim 0.1$, which occupy about $\sim 1$\% of the downstream volume. 
That a significant fraction of the emission is contributed by the high field regions is further supported by the downstream magnetic energy distribution 
(inset of Fig.~\ref{fig:sync}, showing $\varepsilon_B^2f(\varepsilon_B)$), which peaks near $\varepsilon_B\sim 0.1$.

Our results motivate the design of two-zone synchrotron emission models \citep[e.g.,][]{KumarP2012,Khangulyan2021}, where 
particles spend most of their time in low field regions, 
but emit an order unity fraction of their synchrotron power intermittently 
at the locations with strong fields.\footnote{The high field regions in our simulation amount to a small 
fraction of the downstream volume but contain roughly half of the total magnetic energy, which differs from the model 
by \citet{KumarP2012} where the fraction of energy in high field regions is small. The scenario implied by our simulations is closer 
to the model proposed by \citet{Khangulyan2021}.} In the scenario implied by our simulations, 
the average synchrotron cooling rate (for a given $\gamma$) is determined by the mean value $\langle\varepsilon_B\rangle \ll 0.1$ of the 
magnetic energy fraction in the emission region  (see Fig.~\ref{fig:sync}). In contrast, the photon energy 
$E_{\rm syn} \sim \gamma^2\hbar eB_*/m_{\rm e}c$ at the peak of the emitted spectral energy distribution 
is set by the high field strength $B_*$, corresponding to the peak of $\varepsilon_B^2f(\varepsilon_B)$ 
at $\varepsilon_{B*} \sim 0.1$ (inset of Fig.~\ref{fig:sync}), since it is the value at the peak of $\varepsilon_B^2f(\varepsilon_B)$ that makes the single 
largest contribution to the emission. A convenient analytic estimate of $\varepsilon_{B*}$ 
can be given based on the synchrotron-power-weighted mean of $\varepsilon_B$, which equals $\langle\varepsilon_B^2\rangle / \langle\varepsilon_B\rangle$ (for fixed $\gamma$).
In our simulation, indeed $\varepsilon_{B*} \sim \langle\varepsilon_B^2\rangle / \langle\varepsilon_B\rangle \sim 0.1$. Using our analytic estimate,
we can express $E_{\rm syn} \sim \kappa^{1/2} \gamma^2 \hbar e B_{\rm rms} / m_{\rm e} c$, where 
$\kappa = \langle \varepsilon_B^2\rangle / \langle \varepsilon_B\rangle^2$ is 
the kurtosis of the magnetic fluctuations and $B_{\rm rms}\propto\langle\varepsilon_B\rangle^{1/2}$ is the root-mean-square field strength. 
The downstream field at the end of our simulation has $\kappa \approx 25$ on average, which is much greater than the value $\kappa = 3$ for normal Gaussian distributions. 
As a result, $E_{\rm syn}$ is raised by a factor of $\sim 3$ compared to the emission in a non-intermittent field with $\kappa=3$. 
The factor could be increased further for more intermittent distributions with even larger $\kappa$. Note that the factor $\sim\kappa^{1/2}$ 
applies to all the emitting particles, including the highest energy ones. If the highest-energy particles are accelerated at maximum efficiency (i.e., close to
the Bohm regime with $\lambda\sim R_{\rm L}(\gamma)$), the intermittency of the post-shock magnetic field enables the production of synchrotron photons beyond the 
classical burn-off limit \citep{deJager1996,KumarP2012,Khangulyan2021}.

\section{Particle acceleration}
\label{sec:acc}

\begin{figure}[htb!]
\centering
\includegraphics[width=\columnwidth]{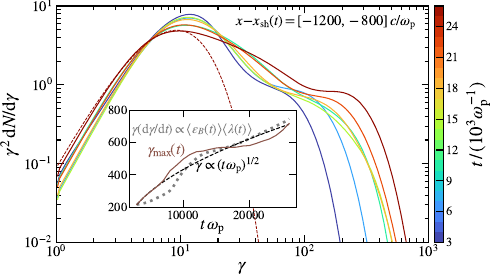}
\caption{\label{fig:spec} Evolution of the particle energy spectrum in a fixed slab behind the shock (colors represent time). The dashed 
red curve shows a Maxwellian fit to the low-energy part of the spectrum at the end of the simulation. The inset shows the evolution 
of the maximum Lorentz factor $\gamma_{\max}(t)$ (brown curve), and compares the measurement with 
theoretical expectations for particle scattering in a magnetic field with fixed $\langle\varepsilon_B\rangle\langle\lambda\omega_{\rm p}/c\rangle\approx 0.4$  (black dashed curve) versus 
scattering in a field where $\langle\varepsilon_B\rangle\langle\lambda\omega_{\rm p}/c\rangle$ is time-evolving  (gray dotted curve; see text for details). The maximum Lorentz factor
is determined by the value of $\gamma$ where $\gamma{\rm d}N/{\rm d}\gamma$ drops by $10^5$ below the peak.}
\end{figure}

Finally, we study particle acceleration at the relativistic shock.
Fig.~\ref{fig:spec} shows the evolution of the particle energy 
spectrum ($\gamma^2{\rm d}N/{\rm d}\gamma$) in a fixed slab behind the shock at $x\!-\!x_{\rm sh}= [-1200,\,-800]\, c/\omega_{\rm p}$. Particles are accelerated to 
higher energies as the simulation progresses, consistent with earlier 
works \citep[e.g.,][]{Spitkovsky2008}. At select times, which correlate 
with bursts of plasma cavity generation (around $t\omega_{\rm p}\sim 10^4$ and toward 
the end of the simulation; see Fig.~\ref{fig:evolution}), 
the particle maximum Lorentz factor in the slab grows 
faster than the canonical $\gamma_{\max}(t) \propto (t\omega_{\rm p})^{1/2}$ scaling \citep{Sironi2013,Plotnikov2018},  expected for Weibel-mediated relativistic 
shocks (see inset of Fig.~\ref{fig:spec}). However, in between the intervals of fast acceleration lies a quiescent phase 
with a very slow growth of $\gamma_{\max}(t)$, so that the maximum Lorentz factor still 
traces out a $\sim (t\omega_{\rm p})^{1/2}$ envelope on average. The rapid changes in the instantaneous rate of particle acceleration
can be largely attributed to the evolution of the magnetic field properties, which set the particle
scattering frequency as $\nu_{\rm scatt} \sim \lambda c / R_{\rm L}^2 \sim \varepsilon_B(\lambda\omega_{\rm p}/c)(\gamma/\gamma_0)^{-2}\omega_{\rm p}$
\citep[e.g.,][]{Plotnikov2011}. An integration of ${\rm d}\gamma/{\rm d t} = 0.25\,\nu_{\rm scatt}\gamma$ (using 0.25 as an \emph{ad hoc} prefactor)
with a time-dependent $\langle\varepsilon_B\rangle\langle\lambda\omega_{\rm p}/c\rangle$, measured in the same downstream slab as the particle spectrum (see Fig.~\ref{fig:statistics}),
gives a result broadly consistent with the measured $\gamma_{\max}(t)$ (gray dotted curve in the inset of Fig.~\ref{fig:spec}). A 
reasonable overall agreement is also obtained for a fixed $\langle\varepsilon_B\rangle\langle\lambda\omega_{\rm p}/c\rangle\approx 0.4$ (dashed black curve).

A novel feature observed in the particle spectrum at late times is the formation of a distinct
suprathermal component, which connects the Maxwellian peak (near $\gamma\sim\gamma_0 = 10$)
to the high-energy power law tail with an index close to $-2$ at late time (i.e., ${\rm d}N/{\rm d}\gamma\propto\gamma^{-2}$ at $100\lesssim\gamma\lesssim 200$).
The fraction of kinetic energy contained in the suprathermal component grows with time. Toward the end of the run, the suprathermal component contains 
about 30\% of the total kinetic energy, while the high-energy tail contains stably around 10\%. The suprathermal component grows at the expense of the
thermal (i.e., Maxwellian) part of the distribution, which loses energy over time.
It is interesting to note that the start of the high-energy tail shifts with time to
higher $\gamma$, in favor of the growing suprathermal component. In this regard, let us mention that standard treatments of the first order 
Fermi process at relativistic shocks \citep[e.g.,][]{Achterberg2001} assume a discontinuous 
flow profile across the shock transition. However, in practice the shock develops a broad transition
region over which the incoming flow gradually slows down \citep{Lemoine2019}. 
The enhanced scattering due to large-scale structures and the strongly turbulent shock transition layer observed in our simulation
imply that moderate energy particles ($\gamma\sim$ a few $\gamma_0$) returning into the upstream may deflect toward 
the downstream before being able to probe the full velocity difference between the far upstream and downstream.
We speculate that these particles represent the dominant contribution to the suprathermal component, but we leave a more 
detailed investigation of this aspect for the future.

\section{Discussion and Conclusions}
\label{sec:discussion}

We studied the evolution of a relativistic pair plasma shock propagating into an unmagnetized medium using 2D PIC simulations 
of unprecedented duration and size. Toward the end of our fiducial run, at $t\omega_{\rm p}\approx 26100$, 
the shock generates magnetic structures reaching sizes of roughly $\sim$ 100 plasma skin depths on both sides of the shock. The size of magnetic structures generated in our simulation may be sufficient 
to circumvent at least some of the outstanding issues related to the modeling of GRB afterglows.
For instance, it has been argued \citep{Huang2022} that magnetic fields with a coherence scale of about 
$\sim 100$ ion skin depths or larger are required to explain the X-ray and gamma-ray 
afterglow of GRB 190829A \citep{HESS2021}.\footnote{For application to GRB shocks propagating into an electron-proton ambient medium, 
time and length scales should be measured in units of the ion (i.e., proton) inverse plasma frequency $\omega_{\rm pi}^{-1}= (m_{\rm i}/m_{\rm e})^{1/2}\omega_{\rm p}^{-1}$ and 
skin depth $c/\omega_{\rm pi} = (m_{\rm i}/m_{\rm e})^{1/2} c/\omega_{\rm p}$, where $m_{\rm i}/m_{\rm e}$ is the ion-electron mass ratio. In our simulation, 
$m_{\rm i}/m_{\rm e}=1$ for computational convenience.}
It should be noted that further evolution of the shock structure 
is expected beyond the time span of our simulation, as discussed below. 
We also find that the bulk of the post-shock plasma becomes magnetized,
which slows down the magnetic field decay, as compared to early times of the shock evolution (Sec.~\ref{sec:stat}).
The probability distribution of the downstream magnetic energy fraction $\varepsilon_B$ 
is intermittent; it features a hard power-law tail which extends up to $\varepsilon_B \sim 0.1$. Particles 
randomly sampling the downstream field spend most of their time in low field regions ($\varepsilon_B \ll 0.1$),
but emit roughly half of their synchrotron power at localized magnetic structures with $\varepsilon_B \sim 0.1$.

The shock exhibits rich spatiotemporal dynamics, characterized by bursts 
of magnetized plasma cavity generation and oscillations in the sign of their polarity (Sec.~\ref{sec:evolution}). The cavitation instability enables the
formation of large-scale magnetic structures 
over the turbulent shock precursor, giving 
rise to a broadband 
magnetic wavenumber spectrum at the end of the simulation (Fig.~\hyperref[fig:statistics]{\ref*{fig:statistics}(f)}).
The spectrum extends from the large energy-containing scale, characterized by the transverse coherence scale $\lambda$,
down to plasma microscales $\sim c/\omega_{\rm p}$. Empirically, we find that $\lambda$ measured immediately ahead of the shock is proportional to $\Delta x_{\rm cav}$,
where $\Delta x_{\rm cav}$ is the distance between the shock and the upstream location of cavity birth (Fig.~\hyperref[fig:statistics]{\ref*{fig:statistics}(d)}). Beyond the duration of our simulation,
an upper limit on $\lambda(t)$ may be obtained by assuming that the distance $\Delta x_{\rm cav}$ scales linearly with the size of the 
precursor $\ell_{\rm prec}$, which is set by the scattering length of the highest energy particles. The upstream scattering length 
$\ell_{\rm scatt}\propto \gamma^2$ \citep{Lemoine2019c}. This implies $\lambda \propto t$, assuming the shock keeps 
accelerating particles to higher energy as $\gamma_{\rm max}(t)\propto t^{1/2}$ (see Sec.~\ref{sec:acc}). A linear growth rate for $\lambda$ 
presents a potentially favorable scenario for particle acceleration; it suggests that the ratio 
$\lambda/R_{\rm L}(\gamma_{\rm max})\sim \varepsilon_B^{1/2}(\lambda\omega_{\rm p}/c)(\gamma_{\rm max}/\gamma_0)^{-1}$ grows over time 
and converges toward the Bohm limit ($\lambda \sim R_{\rm L}(\gamma_{\max})$).
While these results are encouraging, we 
caution the reader that extrapolations beyond the time span of our simulation are uncertain, and the 
scaling $\lambda\propto t$ is only an optimistic estimate.

The size of magnetic structures increases from early to late time also behind the shock front (Fig.~\ref{fig:structure}).
However, quantitatively the increase of the field coherence scale $\lambda$ in the downstream is not large \citep[see also][]{Keshet2009}. 
At different times and/or at different locations behind the shock, we measure 
typical values of the order of $\lambda\sim 100\, c/\omega_{\rm p}$ (Fig.~\hyperref[fig:statistics]{\ref*{fig:statistics}(c)}). This 
scale corresponds within factors of a few to the wavenumber $k_y\sim 0.1\,\omega_{\rm p}/c$ at which the magnetic spectrum peaks 
(Fig.~\hyperref[fig:statistics]{\ref*{fig:statistics}(e)}), which also happens to be the scale at which 
the bulk of the post-shock plasma becomes marginally magnetized. 
Presumably, the demonstration of a large increase of $\lambda$ in the downstream requires longer duration simulations 
with even wider boxes. In particular, we expect the coherence scale at a fixed distance behind the shock to grow when the size of magnetic structures arriving at 
the shock exceeds our ``typical'' downstream value of $\sim 100\, c/\omega_{\rm p}$. Then, as the growing magnetic structures are transmitted into the downstream, the 
coherence scale will increase on both sides of the shock.

Even though the presented PIC simulation is the largest and 
longest of its kind, extrapolations are still required in order 
to make direct contact with GRB afterglows. The apparent duration of
our simulation in the observer frame \citep{Meszaros2006}
is $T_{\rm obs}\simeq T(1+z)/2\gamma_0\approx 0.02\,(\gamma_0/10)^{-1}(1+z)(n_{\rm u}/1\,{\rm cm}^{-3})^{-1/2}(m_{\rm i}/m_{\rm e})^{1/2}\,{\rm s}$, where $T$ is the duration in the (downstream) simulation frame, 
$n_{\rm u}$ is the upstream-frame number
density of the ambient medium, $z$ is the redshift, and $m_{\rm i}/m_{\rm e}$ is the ion-electron mass ratio ($m_{\rm i}/m_{\rm e}= 1$ in our run, but for electron-proton shocks $m_{\rm i}/m_{\rm e}\approx 1836$). The
transverse size of the box is
$L_\perp\approx 7\times 10^8(n_{\rm u}/1\,{\rm cm}^{-3})^{-1/2}(m_{\rm i}/m_{\rm e})^{1/2}\,{\rm cm}$. 
Thus, our
present simulation probes the shock physics over time and length scales 
intermediate between the plasma microscales and the global macroscales of the blast wave.
In our fiducial run, spatial regions featuring cavities of a given polarity extend up to thousands of skin depths in the longitudinal direction, and across the full transverse dimension of the simulation box. In a macroscopically large
domain, these long-range correlations will be ultimately 
limited by causality. In that case, 
we expect the shock to form cavities of different polarity in causally disconnected regions transverse 
to the shock normal. Our present simulation box is too narrow to cancel out the system-wide correlations. Strictly speaking, the exact times and locations of cavity generation observed in our fiducial run 
represent only a particular realization of our numerical experiment. Nevertheless, the general trend of large-scale field generation
is still clearly evident when considering the full time span of the simulation.

Previous PIC simulations found that as much as $\sim 90$\% of the post-shock kinetic 
energy is contained in thermal (i.e., Maxwellian) particles \citep[e.g.,][]{Spitkovsky2008,Spitkovsky2008b,Martins2009}. 
Radiative signatures of thermal electrons from GRB afterglows have been considered by
a number of authors \citep[e.g.,][]{Eichler2005,Dimitrios2009,Ressler2017,Laskar2019,Warren2022}. However, conclusive 
observational evidence for the presence of a prominent thermal component is presently lacking. In contrast to previous PIC simulations, 
we find that the fraction of energy contained in the thermal component drops over time. At the end of our fiducial run, the fraction
of energy in the Maxwellian component drops from the ``canonical'' $\sim 90$\% to $\sim 60$\% (Sec.~\ref{sec:acc}), which reduces
the tension with the widely used assumption of purely nonthermal electrons in afterglow models \citep[e.g.,][]{Sari1998,Granot2002}.

In order to evolve the shock simulations for a maximum possible duration we focused on a pair plasma composition. Previous works 
showed that the physics of electron-ion relativistic unmagnetized shocks is similar to pair shocks, owing to efficient preheating
of the incoming electrons \citep[e.g.,][]{Spitkovsky2008b,Martins2009,Sironi2013}. Therefore, we expect qualitatively similar results for electron-ion 
shocks. A critical parameter affecting the shock evolution is the 
upstream magnetization of the ambient medium $\sigma$ \citep[see][]{Sironi2013, Reville2014, Plotnikov2018}. The physical picture described here
requires that the shock operates close to the $\sigma = 0$ limit. Presently, it is not well understood how small $\sigma$ 
needs to be for the $\sigma=0$ limit to be relevant. 
It has been shown that different levels of ambient magnetization can dramatically alter the behavior 
of collisionless shocks \citep[e.g.,][]{Bret2017,Bret2018,Haggerty2022,Groselj2022}. 
Thus, it may be reasonably expected that only a fraction of GRBs take
place in environments where a vanishingly small magnetization can be assumed, which might help explaining the relatively large 
scatter of best-fit modeling parameters with respect to different afterglow observations.

\begin{acknowledgments}
We thank A.~Vanthieghem, M.~Lemoine, B.~Reville, and A.~\mbox{Beloborodov} for helpful discussions related to this work.
D.G.~is supported by the Research Foundation -- Flanders (FWO) Senior Postdoctoral 
Fellowship 12B1424N. L.S.~and D.G.~were also supported by NASA ATP grant 80NSSC20K0565.
This research was facilitated by Multimessenger Plasma Physics Center (MPPC), NSF grants PHY-2206607 and PHY-2206609. 
The work was supported by a grant from the Simons Foundation (MP-SCMPS-0000147, to L.S.~and A.S.). 
L.S.~acknowledges support from DoE Early Career Award DE-SC0023015.
Simulations were performed on NASA Pleiades (GID: s2796, s1967, s2560). 
We acknowledge the OSIRIS Consortium, consisting of UCLA and IST (Lisbon, Portugal) for the 
use of \textsc{osiris} and for providing access to the \textsc{osiris} framework.
\end{acknowledgments}
{\software{\textsc{osiris} \citep{Fonseca2002,Fonseca2013}}}

\appendix

\section{Measurement of $\Delta x_{\rm cav}$}
\label{sec:symmetry}

We employ the order parameter $Q(x)=\langle{\rm sgn}(-A_x)\rangle_{\!y}$ (see Fig.~\ref{fig:evolution}) to determine the typical distance $\Delta x_{\rm cav}$ 
between the upstream location where the cavities are born and the shock. The distance $\Delta x_{\rm cav}$
determines the size of the cavities arriving at the shock (see Sec.~\ref{sec:stat}). We implemented an
algorithm to measure $\Delta x_{\rm cav}$ as follows. 
To reduce noise we first smooth $Q(x)$ over a longitudinal scale of 100 $c/\omega_{\rm p}$ and we set
$Q(x) = 0$ where $\langle\varepsilon_B\rangle_{\!y} < 10^{-6}$, since $A_x$ cannot be reliably 
calculated when the magnetic fluctuations are vanishingly small. As an \emph{ad hoc} condition for 
statistical significance we introduce $\max|Q(x\!\!>\!\!x_{\rm sh})| > \epsilon$ with $\epsilon = 0.1$. At $t\omega_{\rm p}\lesssim 4000$ 
and during a quiescent phase around $t\omega_{\rm p}\sim 12000$ cavities are largely absent 
and $\max|Q(x\!\!>\!\!x_{\rm sh})| < \epsilon$, so that $\Delta x_{\rm cav}$ is not measured.
If the condition for statistical significance is satisfied we proceed with the calculation.
Spatially separated regions of the precursor can simultaneously generate cavities of different polarity (e.g., see Fig.~\hyperref[fig:evolution]{\ref*{fig:evolution}(d)} and the
corresponding profile of $Q(x)$),
but it is the cavities that make the largest contribution to the symmetry breaking, which dominate the scale of magnetic structures ultimately arriving at the shock.
For this reason, we first determine the polarity of the dominant cavities as $P = {\rm sgn}\langle Q(x\!\!>\!\!x_{\rm sh})\rangle_{\!x}$. 
We then measure $\Delta x_{\rm cav}$ as the largest distance between $x_{\rm sh}$ and any upstream location with $PQ(x) > \epsilon/2$. 
We visually inspected the magnetic field snapshots and confirmed that our method gives 
reasonable estimates for the upstream location $x_{\rm cav} = x_{\rm sh} + \Delta x_{\rm cav}$ of cavity birth.
For reference, we show the measured locations with 
squares in Fig.~\hyperref[fig:evolution]{\ref*{fig:evolution}(f)} (measurements are absent for $t\omega_{\rm p} = 3000,\, 13400$ because cavities are not present 
at a statistically significant level at those times).

\bibliographystyle{aasjournal}

\begin{thebibliography}{}
\expandafter\ifx\csname natexlab\endcsname\relax\def\natexlab#1{#1}\fi
\providecommand{\url}[1]{\href{#1}{#1}}
\providecommand{\dodoi}[1]{doi:~\href{http://doi.org/#1}{\nolinkurl{#1}}}
\providecommand{\doeprint}[1]{\href{http://ascl.net/#1}{\nolinkurl{http://ascl.net/#1}}}
\providecommand{\doarXiv}[1]{\href{https://arxiv.org/abs/#1}{\nolinkurl{https://arxiv.org/abs/#1}}}

\bibitem[{{Abdalla} {et~al.}(2019){Abdalla}, {Adam}, {Aharonian}, {Ait
  Benkhali}, {Ang{\"u}ner}, {Arakawa}, {Arcaro}, {Armand}, {Ashkar}, {Backes},
  {Barbosa Martins}, {Barnard}, {Becherini}, {Berge}, {Bernl{\"o}hr},
  {Bissaldi}, {Blackwell}, {B{\"o}ttcher}, {Boisson}, {Bolmont}, {Bonnefoy},
  {Bregeon}, {Breuhaus}, {Brun}, {Brun}, {Bryan}, {B{\"u}chele}, {Bulik},
  {Bylund}, {Capasso}, {Caroff}, {Carosi}, {Casanova}, {Cerruti}, {Chand},
  {Chandra}, {Chen}, {Colafrancesco}, {Cury{\l}o}, {Davids}, {Deil}, {Devin},
  {deWilt}, {Dirson}, {Djannati-Ata{\"\i}}, {Dmytriiev}, {Donath},
  {Doroshenko}, {Dyks}, {Egberts}, {Emery}, {Ernenwein}, {Eschbach}, {Feijen},
  {Fegan}, {Fiasson}, {Fontaine}, {Funk}, {F{\"u}{\ss}ling}, {Gabici},
  {Gallant}, {Gat{\'e}}, {Giavitto}, {Giunti}, {Glawion}, {Glicenstein},
  {Gottschall}, {Grondin}, {Hahn}, {Haupt}, {Heinzelmann}, {Henri}, {Hermann},
  {Hinton}, {Hofmann}, {Hoischen}, {Holch}, {Holler}, {Horns}, {Huber},
  {Iwasaki}, {Jamrozy}, {Jankowsky}, {Jankowsky}, {Jardin-Blicq},
  {Jung-Richardt}, {Kastendieck}, {Katarzy{\'n}ski}, {Katsuragawa}, {Katz},
  {Khangulyan}, {Kh{\'e}lifi}, {King}, {Klepser}, {Klu{\'z}niak}, {Komin},
  {Kosack}, {Kostunin}, {Kreter}, {Lamanna}, {Lemi{\`e}re}, {Lemoine-Goumard},
  {Lenain}, {Leser}, {Levy}, {Lohse}, {Lypova}, {Mackey}, {Majumdar},
  {Malyshev}, {Marandon}, {Marcowith}, {Mares}, {Mariaud}, {Mart{\'\i}-Devesa},
  {Marx}, {Maurin}, {Meintjes}, {Mitchell}, {Moderski}, {Mohamed}, {Mohrmann},
  {Moore}, {Moulin}, {Muller}, {Murach}, {Nakashima}, {de Naurois},
  {Ndiyavala}, {Niederwanger}, {Niemiec}, {Oakes}, {O'Brien}, {Odaka}, {Ohm},
  {de Ona Wilhelmi}, {Ostrowski}, {Oya}, {Panter}, {Parsons}, {Perennes},
  {Petrucci}, {Peyaud}, {Piel}, {Pita}, {Poireau}, {Priyana Noel}, {Prokhorov},
  {Prokoph}, {P{\"u}hlhofer}, {Punch}, {Quirrenbach}, {Raab}, {Rauth},
  {Reimer}, {Reimer}, {Remy}, {Renaud}, {Rieger}, {Rinchiuso}, {Romoli},
  {Rowell}, {Rudak}, {Ruiz-Velasco}, {Sahakian}, {Sailer}, {Saito}, {Sanchez},
  {Santangelo}, {Sasaki}, {Schlickeiser}, {Sch{\"u}ssler}, {Schulz}, {Schutte},
  {Schwanke}, {Schwemmer}, {Seglar-Arroyo}, {Senniappan}, {Seyffert}, {Shafi},
  {Shiningayamwe}, {Simoni}, {Sinha}, {Sol}, {Specovius}, {Spir-Jacob},
  {Stawarz}, {Steenkamp}, {Stegmann}, {Steppa}, {Takahashi}, {Tavernier},
  {Taylor}, {Terrier}, {Tiziani}, {Tluczykont}, {Trichard}, {Tsirou}, {Tsuji},
  {Tuffs}, {Uchiyama}, {van der Walt}, {van Eldik}, {van Rensburg}, {van
  Soelen}, {Vasileiadis}, {Veh}, {Venter}, {Vincent}, {Vink}, {V{\"o}lk},
  {Vuillaume}, {Wadiasingh}, {Wagner}, {White}, {Wierzcholska}, {Yang},
  {Yoneda}, {Zacharias}, {Zanin}, {Zdziarski}, {Zech}, {Ziegler}, {Zorn},
  {{\.Z}ywucka}, {de Palma}, {Axelsson}, \& {Roberts}}]{HESS2019}
{Abdalla}, H., {Adam}, R., {Aharonian}, F., {et~al.} 2019, \nat, 575, 464,
  \dodoi{10.1038/s41586-019-1743-9}

\bibitem[{{Abdo} {et~al.}(2009){Abdo}, {Ackermann}, {Arimoto}, {Asano},
  {Atwood}, {Axelsson}, {Baldini}, {Ballet}, {Band}, {Barbiellini}, {Baring},
  {Bastieri}, {Battelino}, {Baughman}, {Bechtol}, {Bellardi}, {Bellazzini},
  {Berenji}, {Bhat}, {Bissaldi}, {Blandford}, {Bloom}, {Bogaert}, {Bogart},
  {Bonamente}, {Bonnell}, {Borgland}, {Bouvier}, {Bregeon}, {Brez}, {Briggs},
  {Brigida}, {Bruel}, {Burnett}, {Burrows}, {Busetto}, {Caliandro}, {Cameron},
  {Caraveo}, {Casandjian}, {Ceccanti}, {Cecchi}, {Celotti}, {Charles},
  {Chekhtman}, {Cheung}, {Chiang}, {Ciprini}, {Claus}, {Cohen-Tanugi},
  {Cominsky}, {Connaughton}, {Conrad}, {Costamante}, {Cutini}, {DeKlotz},
  {Dermer}, {de Angelis}, {de Palma}, {Digel}, {Dingus}, {do Couto e Silva},
  {Drell}, {Dubois}, {Dumora}, {Edmonds}, {Evans}, {Fabiani}, {Farnier},
  {Favuzzi}, {Finke}, {Fishman}, {Focke}, {Frailis}, {Fukazawa}, {Funk},
  {Fusco}, {Gargano}, {Gasparrini}, {Gehrels}, {Germani}, {Giebels},
  {Giglietto}, {Giommi}, {Giordano}, {Glanzman}, {Godfrey}, {Goldstein},
  {Granot}, {Greiner}, {Grenier}, {Grondin}, {Grove}, {Guillemot}, {Guiriec},
  {Haller}, {Hanabata}, {Harding}, {Hayashida}, {Hays}, {Morata}, {Hoover},
  {Hughes}, {J{\'o}hannesson}, {Johnson}, {Johnson}, {Johnson}, {Johnson},
  {Kamae}, {Katagiri}, {Kataoka}, {Kavelaars}, {Kawai}, {Kelly}, {Kennea},
  {Kerr}, {Kippen}, {Kn{\"o}dlseder}, {Kocevski}, {Kocian}, {Komin},
  {Kouveliotou}, {Kuehn}, {Kuss}, {Lande}, {Landriu}, {Larsson}, {Latronico},
  {Lavalley}, {Lee}, {Lee}, {Lemoine-Goumard}, {Lichti}, {Longo}, {Loparco},
  {Lott}, {Lovellette}, {Lubrano}, {Madejski}, {Makeev}, {Marangelli},
  {Mazziotta}, {McBreen}, {McEnery}, {McGlynn}, {Meegan}, {M{\'e}sz{\'a}ros},
  {Meurer}, {Michelson}, {Minuti}, {Mirizzi}, {Mitthumsiri}, {Mizuno},
  {Moiseev}, {Monte}, {Monzani}, {Moretti}, {Morselli}, {Moskalenko}, {Murgia},
  {Nakamori}, {Nelson}, {Nolan}, {Norris}, {Nuss}, {Ohno}, {Ohsugi}, {Okumura},
  {Omodei}, {Orlando}, {Ormes}, {Ozaki}, {Paciesas}, {Paneque}, {Panetta},
  {Parent}, {Pelassa}, {Pepe}, {Perri}, {Pesce-Rollins}, {Petrosian},
  {Pinchera}, {Piron}, {Porter}, {Preece}, {Rain{\`o}}, {Ramirez-Ruiz},
  {Rando}, {Rapposelli}, {Razzano}, {Razzaque}, {Rea}, {Reimer}, {Reimer},
  {Reposeur}, {Reyes}, {Ritz}, {Rochester}, {Rodriguez}, {Roth}, {Ryde},
  {Sadrozinski}, {Sanchez}, {Sander}, {Parkinson}, {Scargle}, {Schalk},
  {Segal}, {Sgr{\`o}}, {Shimokawabe}, {Siskind}, {Smith}, {Smith}, {Spandre},
  {Spinelli}, {Stamatikos}, {Starck}, {Stecker}, {Steinle}, {Stephens},
  {Strickman}, {Suson}, {Tagliaferri}, {Tajima}, {Takahashi}, {Takahashi},
  {Tanaka}, {Tenze}, {Thayer}, {Thayer}, {Thompson}, {Tibaldo}, {Torres},
  {Tosti}, {Tramacere}, {Turri}, {Tuvi}, {Usher}, {van der Horst}, {Vigiani},
  {Vilchez}, {Vitale}, {von Kienlin}, {Waite}, {Williams}, {Wilson-Hodge},
  {Winer}, {Wood}, {Wu}, {Yamazaki}, {Ylinen}, {Ziegler}, {Fermi LAT
  Collaboration}, \& {Fermi GBM Collaboration}}]{Abdo2009}
{Abdo}, A.~A., {Ackermann}, M., {Arimoto}, M., {et~al.} 2009, Science, 323,
  1688, \dodoi{10.1126/science.1169101}

\bibitem[{{Achterberg} {et~al.}(2001){Achterberg}, {Gallant}, {Kirk}, \&
  {Guthmann}}]{Achterberg2001}
{Achterberg}, A., {Gallant}, Y.~A., {Kirk}, J.~G., \& {Guthmann}, A.~W. 2001,
  \mnras, 328, 393, \dodoi{10.1046/j.1365-8711.2001.04851.x}

\bibitem[{{Achterberg} {et~al.}(2007){Achterberg}, {Wiersma}, \&
  {Norman}}]{Achterberg2007b}
{Achterberg}, A., {Wiersma}, J., \& {Norman}, C.~A. 2007, \aap, 475, 19,
  \dodoi{10.1051/0004-6361:20065366}

\bibitem[{{Ackermann} {et~al.}(2010){Ackermann}, {Asano}, {Atwood}, {Axelsson},
  {Baldini}, {Ballet}, {Barbiellini}, {Baring}, {Bastieri}, {Bechtol},
  {Bellazzini}, {Berenji}, {Bhat}, {Bissaldi}, {Blandford}, {Bloom},
  {Bonamente}, {Borgland}, {Bouvier}, {Bregeon}, {Brez}, {Briggs}, {Brigida},
  {Bruel}, {Buson}, {Caliandro}, {Cameron}, {Caraveo}, {Carrigan},
  {Casandjian}, {Cecchi}, {{\c{C}}elik}, {Charles}, {Chiang}, {Ciprini},
  {Claus}, {Cohen-Tanugi}, {Connaughton}, {Conrad}, {Dermer}, {de Palma},
  {Dingus}, {Silva}, {Drell}, {Dubois}, {Dumora}, {Farnier}, {Favuzzi},
  {Fegan}, {Finke}, {Focke}, {Frailis}, {Fukazawa}, {Fusco}, {Gargano},
  {Gasparrini}, {Gehrels}, {Germani}, {Giglietto}, {Giordano}, {Glanzman},
  {Godfrey}, {Granot}, {Grenier}, {Grondin}, {Grove}, {Guiriec}, {Hadasch},
  {Harding}, {Hays}, {Horan}, {Hughes}, {J{\'o}hannesson}, {Johnson}, {Kamae},
  {Katagiri}, {Kataoka}, {Kawai}, {Kippen}, {Kn{\"o}dlseder}, {Kocevski},
  {Kouveliotou}, {Kuss}, {Lande}, {Latronico}, {Lemoine-Goumard}, {Llena
  Garde}, {Longo}, {Loparco}, {Lott}, {Lovellette}, {Lubrano}, {Makeev},
  {Mazziotta}, {McEnery}, {McGlynn}, {Meegan}, {M{\'e}sz{\'a}ros}, {Michelson},
  {Mitthumsiri}, {Mizuno}, {Moiseev}, {Monte}, {Monzani}, {Moretti},
  {Morselli}, {Moskalenko}, {Murgia}, {Nakajima}, {Nakamori}, {Nolan},
  {Norris}, {Nuss}, {Ohno}, {Ohsugi}, {Omodei}, {Orlando}, {Ormes}, {Ozaki},
  {Paciesas}, {Paneque}, {Panetta}, {Parent}, {Pelassa}, {Pepe},
  {Pesce-Rollins}, {Piron}, {Preece}, {Rain{\`o}}, {Rando}, {Razzano},
  {Razzaque}, {Reimer}, {Ritz}, {Rodriguez}, {Roth}, {Ryde}, {Sadrozinski},
  {Sander}, {Scargle}, {Schalk}, {Sgr{\`o}}, {Siskind}, {Smith}, {Spandre},
  {Spinelli}, {Stamatikos}, {Stecker}, {Strickman}, {Suson}, {Tajima},
  {Takahashi}, {Takahashi}, {Tanaka}, {Thayer}, {Thayer}, {Thompson},
  {Tibaldo}, {Toma}, {Torres}, {Tosti}, {Tramacere}, {Uchiyama}, {Uehara},
  {Usher}, {van der Horst}, {Vasileiou}, {Vilchez}, {Vitale}, {von Kienlin},
  {Waite}, {Wang}, {Wilson-Hodge}, {Winer}, {Wu}, {Yamazaki}, {Yang}, {Ylinen},
  \& {Ziegler}}]{Ackermann2010}
{Ackermann}, M., {Asano}, K., {Atwood}, W.~B., {et~al.} 2010, \apj, 716, 1178,
  \dodoi{10.1088/0004-637X/716/2/1178}

\bibitem[{{Ackermann} {et~al.}(2014){Ackermann}, {Ajello}, {Asano}, {Atwood},
  {Axelsson}, {Baldini}, {Ballet}, {Barbiellini}, {Baring}, {Bastieri},
  {Bechtol}, {Bellazzini}, {Bissaldi}, {Bonamente}, {Bregeon}, {Brigida},
  {Bruel}, {Buehler}, {Burgess}, {Buson}, {Caliandro}, {Cameron}, {Caraveo},
  {Cecchi}, {Chaplin}, {Charles}, {Chekhtman}, {Cheung}, {Chiang}, {Chiaro},
  {Ciprini}, {Claus}, {Cleveland}, {Cohen-Tanugi}, {Collazzi}, {Cominsky},
  {Connaughton}, {Conrad}, {Cutini}, {D'Ammando}, {de Angelis}, {DeKlotz}, {de
  Palma}, {Dermer}, {Desiante}, {Diekmann}, {Di Venere}, {Drell},
  {Drlica-Wagner}, {Favuzzi}, {Fegan}, {Ferrara}, {Finke}, {Fitzpatrick},
  {Focke}, {Franckowiak}, {Fukazawa}, {Funk}, {Fusco}, {Gargano}, {Gehrels},
  {Germani}, {Gibby}, {Giglietto}, {Giles}, {Giordano}, {Giroletti}, {Godfrey},
  {Granot}, {Grenier}, {Grove}, {Gruber}, {Guiriec}, {Hadasch}, {Hanabata},
  {Harding}, {Hayashida}, {Hays}, {Horan}, {Hughes}, {Inoue}, {Jogler},
  {J{\'o}hannesson}, {Johnson}, {Kawano}, {Kn{\"o}dlseder}, {Kocevski}, {Kuss},
  {Lande}, {Larsson}, {Latronico}, {Longo}, {Loparco}, {Lovellette}, {Lubrano},
  {Mayer}, {Mazziotta}, {McEnery}, {Michelson}, {Mizuno}, {Moiseev}, {Monzani},
  {Moretti}, {Morselli}, {Moskalenko}, {Murgia}, {Nemmen}, {Nuss}, {Ohno},
  {Ohsugi}, {Okumura}, {Omodei}, {Orienti}, {Paneque}, {Pelassa}, {Perkins},
  {Pesce-Rollins}, {Petrosian}, {Piron}, {Pivato}, {Porter}, {Racusin},
  {Rain{\`o}}, {Rando}, {Razzano}, {Razzaque}, {Reimer}, {Reimer}, {Ritz},
  {Roth}, {Ryde}, {Sartori}, {Parkinson}, {Scargle}, {Schulz}, {Sgr{\`o}},
  {Siskind}, {Sonbas}, {Spandre}, {Spinelli}, {Tajima}, {Takahashi}, {Thayer},
  {Thayer}, {Thompson}, {Tibaldo}, {Tinivella}, {Torres}, {Tosti}, {Troja},
  {Usher}, {Vandenbroucke}, {Vasileiou}, {Vianello}, {Vitale}, {Winer}, {Wood},
  {Yamazaki}, {Younes}, {Yu}, {Zhu}, {Bhat}, {Briggs}, {Byrne}, {Foley},
  {Goldstein}, {Jenke}, {Kippen}, {Kouveliotou}, {McBreen}, {Meegan},
  {Paciesas}, {Preece}, {Rau}, {Tierney}, {van der Horst}, {von Kienlin},
  {Wilson-Hodge}, {Xiong}, {Cusumano}, {La Parola}, \&
  {Cummings}}]{Ackermann2014}
{Ackermann}, M., {Ajello}, M., {Asano}, K., {et~al.} 2014, Science, 343, 42,
  \dodoi{10.1126/science.1242353}

\bibitem[{{Asano} {et~al.}(2020){Asano}, {Murase}, \& {Toma}}]{Asano2020}
{Asano}, K., {Murase}, K., \& {Toma}, K. 2020, \apj, 905, 105,
  \dodoi{10.3847/1538-4357/abc82c}

\bibitem[{{Bell}(1978)}]{Bell1978}
{Bell}, A.~R. 1978, \mnras, 182, 147, \dodoi{10.1093/mnras/182.2.147}

\bibitem[{{Blandford} \& {Eichler}(1987)}]{Blandford1987}
{Blandford}, R., \& {Eichler}, D. 1987, \physrep, 154, 1,
  \dodoi{10.1016/0370-1573(87)90134-7}

\bibitem[{{Blandford} \& {Ostriker}(1978)}]{Blandford1978}
{Blandford}, R.~D., \& {Ostriker}, J.~P. 1978, \apjl, 221, L29,
  \dodoi{10.1086/182658}

\bibitem[{{Blinne} {et~al.}(2018){Blinne}, {Schinkel}, {Kuschel}, {Elkina},
  {Rykovanov}, \& {Zepf}}]{Blinne2018}
{Blinne}, A., {Schinkel}, D., {Kuschel}, S., {et~al.} 2018, Computer Physics
  Communications, 224, 273, \dodoi{10.1016/j.cpc.2017.10.010}

\bibitem[{{Bret} \& {Narayan}(2018)}]{Bret2018}
{Bret}, A., \& {Narayan}, R. 2018, Journal of Plasma Physics, 84, 905840604,
  \dodoi{10.1017/S0022377818001125}

\bibitem[{{Bret} {et~al.}(2017){Bret}, {Pe'Er}, {Sironi}, {S{\k{a}}dowski}, \&
  {Narayan}}]{Bret2017}
{Bret}, A., {Pe'Er}, A., {Sironi}, L., {S{\k{a}}dowski}, A., \& {Narayan}, R.
  2017, Journal of Plasma Physics, 83, 715830201,
  \dodoi{10.1017/S0022377817000290}

\bibitem[{{Bret} {et~al.}(2014){Bret}, {Stockem}, {Narayan}, \&
  {Silva}}]{Bret2014}
{Bret}, A., {Stockem}, A., {Narayan}, R., \& {Silva}, L.~O. 2014, Physics of
  Plasmas, 21, 072301, \dodoi{10.1063/1.4886121}

\bibitem[{{Burns} {et~al.}(2023){Burns}, {Svinkin}, {Fenimore}, {Kann},
  {Ag{\"u}{\'\i} Fern{\'a}ndez}, {Frederiks}, {Hamburg}, {Lesage}, {Temiraev},
  {Tsvetkova}, {Bissaldi}, {Briggs}, {Dalessi}, {Dunwoody}, {Fletcher},
  {Goldstein}, {Hui}, {Hristov}, {Kocevski}, {Lysenko}, {Mailyan}, {Mangan},
  {McBreen}, {Racusin}, {Ridnaia}, {Roberts}, {Ulanov}, {Veres},
  {Wilson-Hodge}, \& {Wood}}]{Burns2023}
{Burns}, E., {Svinkin}, D., {Fenimore}, E., {et~al.} 2023, \apjl, 946, L31,
  \dodoi{10.3847/2041-8213/acc39c}

\bibitem[{{Chandra} \& {Frail}(2012)}]{Chandra2012}
{Chandra}, P., \& {Frail}, D.~A. 2012, \apj, 746, 156,
  \dodoi{10.1088/0004-637X/746/2/156}

\bibitem[{{Chang} {et~al.}(2008){Chang}, {Spitkovsky}, \& {Arons}}]{Chang2008}
{Chang}, P., {Spitkovsky}, A., \& {Arons}, J. 2008, \apj, 674, 378,
  \dodoi{10.1086/524764}

\bibitem[{{Crowther}(2007)}]{Crowther2007}
{Crowther}, P.~A. 2007, \araa, 45, 177,
  \dodoi{10.1146/annurev.astro.45.051806.110615}

\bibitem[{{de Jager} {et~al.}(1996){de Jager}, {Harding}, {Michelson}, {Nel},
  {Nolan}, {Sreekumar}, \& {Thompson}}]{deJager1996}
{de Jager}, O.~C., {Harding}, A.~K., {Michelson}, P.~F., {et~al.} 1996, \apj,
  457, 253, \dodoi{10.1086/176726}

\bibitem[{{De Pasquale} {et~al.}(2010){De Pasquale}, {Schady}, {Kuin}, {Page},
  {Curran}, {Zane}, {Oates}, {Holland}, {Breeveld}, {Hoversten}, {Chincarini},
  {Grupe}, {Abdo}, {Ackermann}, {Ajello}, {Axelsson}, {Baldini}, {Ballet},
  {Barbiellini}, {Baring}, {Bastieri}, {Bechtol}, {Bellazzini}, {Berenji},
  {Bissaldi}, {Blandford}, {Bloom}, {Bonamente}, {Borgland}, {Bouvier},
  {Bregeon}, {Brez}, {Briggs}, {Brigida}, {Bruel}, {Burnett}, {Buson},
  {Caliandro}, {Cameron}, {Caraveo}, {Carrigan}, {Casandjian}, {Cecchi},
  {{\c{C}}elik}, {Chekhtman}, {Chiang}, {Ciprini}, {Claus}, {Cohen-Tanugi},
  {Connaughton}, {Conrad}, {Dermer}, {de Angelis}, {de Palma}, {Dingus},
  {Silva}, {Drell}, {Dubois}, {Dumora}, {Farnier}, {Favuzzi}, {Fegan},
  {Fishman}, {Focke}, {Frailis}, {Fukazawa}, {Funk}, {Fusco}, {Gargano},
  {Gasparrini}, {Gehrels}, {Germani}, {Giglietto}, {Giordano}, {Glanzman},
  {Godfrey}, {Granot}, {Greiner}, {Grenier}, {Grove}, {Guillemot}, {Guiriec},
  {Harding}, {Hayashida}, {Hays}, {Horan}, {Hughes}, {Jackson},
  {J{\'o}hannesson}, {Johnson}, {Johnson}, {Kamae}, {Katagiri}, {Kataoka},
  {Kawai}, {Kerr}, {Kippen}, {Kn{\"o}dlseder}, {Kocevski}, {Kuss}, {Lande},
  {Latronico}, {Lemoine-Goumard}, {Longo}, {Loparco}, {Lott}, {Lovellette},
  {Lubrano}, {Makeev}, {Mazziotta}, {McEnery}, {McGlynn}, {Meegan},
  {M{\'e}sz{\'a}ros}, {Meurer}, {Michelson}, {Mitthumsiri}, {Mizuno}, {Monte},
  {Monzani}, {Moretti}, {Morselli}, {Moskalenko}, {Murgia}, {Nolan}, {Norris},
  {Nuss}, {Ohno}, {Ohsugi}, {Omodei}, {Orlando}, {Ormes}, {Paciesas},
  {Paneque}, {Panetta}, {Parent}, {Pelassa}, {Pepe}, {Pesce-Rollins}, {Piron},
  {Porter}, {Preece}, {Rain{\`o}}, {Rando}, {Razzano}, {Reimer}, {Reimer},
  {Reposeur}, {Ritz}, {Rochester}, {Rodriguez}, {Roth}, {Ryde}, {Sadrozinski},
  {Sander}, {Saz Parkinson}, {Scargle}, {Schalk}, {Sgr{\`o}}, {Siskind},
  {Smith}, {Spandre}, {Spinelli}, {Stamatikos}, {Starck}, {Stecker},
  {Strickman}, {Suson}, {Tajima}, {Takahashi}, {Tanaka}, {Thayer}, {Thayer},
  {Thompson}, {Tibaldo}, {Toma}, {Torres}, {Tosti}, {Tramacere}, {Uchiyama},
  {Uehara}, {Usher}, {van der Horst}, {Vasileiou}, {Vilchez}, {Vitale}, {von
  Kienlin}, {Waite}, {Wang}, {Winer}, {Wood}, {Wu}, {Yamazaki}, {Ylinen}, \&
  {Ziegler}}]{DePasquale2010}
{De Pasquale}, M., {Schady}, P., {Kuin}, N.~P.~M., {et~al.} 2010, \apjl, 709,
  L146, \dodoi{10.1088/2041-8205/709/2/L146}

\bibitem[{{Drury}(1983)}]{Drury1983}
{Drury}, L.~O. 1983, Reports on Progress in Physics, 46, 973,
  \dodoi{10.1088/0034-4885/46/8/002}

\bibitem[{{Eichler} \& {Waxman}(2005)}]{Eichler2005}
{Eichler}, D., \& {Waxman}, E. 2005, \apj, 627, 861, \dodoi{10.1086/430596}

\bibitem[{{Fonseca} {et~al.}(2013){Fonseca}, {Vieira}, {Fiuza}, {Davidson},
  {Tsung}, {Mori}, \& {Silva}}]{Fonseca2013}
{Fonseca}, R.~A., {Vieira}, J., {Fiuza}, F., {et~al.} 2013, Plasma Physics and
  Controlled Fusion, 55, 124011, \dodoi{10.1088/0741-3335/55/12/124011}

\bibitem[{Fonseca {et~al.}(2002)Fonseca, Silva, Tsung, Decyk, Lu, Ren, Mori,
  Deng, Lee, Katsouleas, \& Adam}]{Fonseca2002}
Fonseca, R.~A., Silva, L.~O., Tsung, F.~S., {et~al.} 2002, Lecture Notes in
  Computer Science, 2331, 342, \dodoi{10.1007/3-540-47789-6_36}

\bibitem[{{Fried}(1959)}]{Fried1959}
{Fried}, B.~D. 1959, Physics of Fluids, 2, 337, \dodoi{10.1063/1.1705933}

\bibitem[{{Giannios} \& {Spitkovsky}(2009)}]{Dimitrios2009}
{Giannios}, D., \& {Spitkovsky}, A. 2009, \mnras, 400, 330,
  \dodoi{10.1111/j.1365-2966.2009.15454.x}

\bibitem[{{Granot} \& {Sari}(2002)}]{Granot2002}
{Granot}, J., \& {Sari}, R. 2002, \apj, 568, 820, \dodoi{10.1086/338966}

\bibitem[{{Gro{\v{s}}elj} {et~al.}(2022){Gro{\v{s}}elj}, {Sironi}, \&
  {Beloborodov}}]{Groselj2022}
{Gro{\v{s}}elj}, D., {Sironi}, L., \& {Beloborodov}, A.~M. 2022, \apj, 933, 74,
  \dodoi{10.3847/1538-4357/ac713e}

\bibitem[{{Gruzinov}(2001)}]{Gruzinov2001}
{Gruzinov}, A. 2001, \apjl, 563, L15, \dodoi{10.1086/324223}

\bibitem[{{H.~E.~S.~S. Collaboration} {et~al.}(2021){H.~E.~S.~S.
  Collaboration}, {Abdalla}, {Aharonian}, {Ait Benkhali}, {Ang{\"u}ner},
  {Arcaro}, {Armand}, {Armstrong}, {Ashkar}, {Backes}, {Baghmanyan}, {Barbosa
  Martins}, {Barnacka}, {Barnard}, {Becherini}, {Berge}, {Bernl{\"o}hr}, {Bi},
  {Bissaldi}, {B{\"o}ttcher}, {Boisson}, {Bolmont}, {de Bony de Lavergne},
  {Breuhaus}, {Brun}, {Brun}, {Bryan}, {B{\"u}chele}, {Bulik}, {Bylund},
  {Caroff}, {Carosi}, {Casanova}, {Chand}, {Chandra}, {Chen}, {Cotter},
  {Cury{\l}o}, {Damascene Mbarubucyeye}, {Davids}, {Davies}, {Deil}, {Devin},
  {Dirson}, {Djannati-Ata{\"\i}}, {Dmytriiev}, {Donath}, {Doroshenko},
  {Dreyer}, {Duffy}, {Dyks}, {Egberts}, {Eichhorn}, {Einecke}, {Emery},
  {Ernenwein}, {Feijen}, {Fegan}, {Fiasson}, {Fichet de Clairfontaine},
  {Fontaine}, {Funk}, {F{\"u}{\ss}ling}, {Gabici}, {Gallant}, {Giavitto},
  {Giunti}, {Glawion}, {Glicenstein}, {Grondin}, {Hahn}, {Haupt}, {Hermann},
  {Hinton}, {Hofmann}, {Hoischen}, {Holch}, {Holler}, {H{\"o}rbe}, {Horns},
  {Huber}, {Jamrozy}, {Jankowsky}, {Jankowsky}, {Jardin-Blicq}, {Joshi},
  {Jung-Richardt}, {Kasai}, {Kastendieck}, {Katarzy{\'n}ski}, {Katz},
  {Khangulyan}, {Kh{\'e}lifi}, {Klepser}, {Klu{\'z}niak}, {Komin}, {Konno},
  {Kosack}, {Kostunin}, {Kreter}, {Lamanna}, {Lemi{\`e}re}, {Lemoine-Goumard},
  {Lenain}, {Leuschner}, {Levy}, {Lohse}, {Lypova}, {Mackey}, {Majumdar},
  {Malyshev}, {Malyshev}, {Marandon}, {Marchegiani}, {Marcowith}, {Mares},
  {Mart{\'\i}-Devesa}, {Marx}, {Maurin}, {Meintjes}, {Meyer}, {Mitchell},
  {Moderski}, {Mohrmann}, {Montanari}, {Moore}, {Morris}, {Moulin}, {Muller},
  {Murach}, {Nakashima}, {Nayerhoda}, {de Naurois}, {Ndiyavala}, {Niemiec},
  {Oakes}, {O'Brien}, {Odaka}, {Ohm}, {Olivera-Nieto}, {de Ona Wilhelmi},
  {Ostrowski}, {Panny}, {Panter}, {Parsons}, {Peron}, {Peyaud}, {Piel}, {Pita},
  {Poireau}, {Priyana Noel}, {Prokhorov}, {Prokoph}, {P{\"u}hlhofer}, {Punch},
  {Quirrenbach}, {Raab}, {Rauth}, {Reichherzer}, {Reimer}, {Reimer}, {Remy},
  {Renaud}, {Rieger}, {Rinchiuso}, {Romoli}, {Rowell}, {Rudak}, {Ruiz-Velasco},
  {Sahakian}, {Sailer}, {Salzmann}, {Sanchez}, {Santangelo}, {Sasaki},
  {Scalici}, {Sch{\"a}fer}, {Sch{\"u}ssler}, {Schutte}, {Schwanke},
  {Seglar-Arroyo}, {Senniappan}, {Seyffert}, {Shafi}, {Shapopi},
  {Shiningayamwe}, {Simoni}, {Sinha}, {Sol}, {Specovius}, {Spencer},
  {Spir-Jacob}, {Stawarz}, {Sun}, {Steenkamp}, {Stegmann}, {Steinmassl},
  {Steppa}, {Takahashi}, {Tam}, {Tavernier}, {Taylor}, {Terrier}, {Thiersen},
  {Tiziani}, {Tluczykont}, {Tomankova}, {Tsirou}, {Tuffs}, {Uchiyama}, {van der
  Walt}, {van Eldik}, {van Rensburg}, {van Soelen}, {Vasileiadis}, {Veh},
  {Venter}, {Vincent}, {Vink}, {V{\"o}lk}, {Wadiasingh}, {Wagner}, {Watson},
  {Werner}, {White}, {Wierzcholska}, {Wong}, {Yusafzai}, {Zacharias}, {Zanin},
  {Zargaryan}, {Zdziarski}, {Zech}, {Zhu}, {Zorn}, {Zouari}, {{\.Z}ywucka},
  {Evans}, \& {Page}}]{HESS2021}
{H.~E.~S.~S. Collaboration}, {Abdalla}, H., {Aharonian}, F., {et~al.} 2021,
  Science, 372, 1081, \dodoi{10.1126/science.abe8560}

\bibitem[{{Haggerty} {et~al.}(2022){Haggerty}, {Bret}, \&
  {Caprioli}}]{Haggerty2022}
{Haggerty}, C.~C., {Bret}, A., \& {Caprioli}, D. 2022, \mnras, 509, 2084,
  \dodoi{10.1093/mnras/stab3110}

\bibitem[{{Haugb{\o}lle}(2011)}]{Haugbolle2011}
{Haugb{\o}lle}, T. 2011, \apjl, 739, L42, \dodoi{10.1088/2041-8205/739/2/L42}

\bibitem[{{Honda} {et~al.}(2000){Honda}, {Meyer-ter-Vehn}, \&
  {Pukhov}}]{Honda2000}
{Honda}, M., {Meyer-ter-Vehn}, J., \& {Pukhov}, A. 2000, Physics of Plasmas, 7,
  1302, \dodoi{10.1063/1.873941}

\bibitem[{{Huang} {et~al.}(2022{\natexlab{a}}){Huang}, {Hu}, {Chen}, {Zha},
  {Liu}, {Yao}, {Cao}, \& {Experiment}}]{Huang2022TeV}
{Huang}, Y., {Hu}, S., {Chen}, S., {et~al.} 2022{\natexlab{a}}, GRB Coordinates
  Network, 32677, 1

\bibitem[{{Huang} {et~al.}(2022{\natexlab{b}}){Huang}, {Kirk}, {Giacinti}, \&
  {Reville}}]{Huang2022}
{Huang}, Z.-Q., {Kirk}, J.~G., {Giacinti}, G., \& {Reville}, B.
  2022{\natexlab{b}}, \apj, 925, 182, \dodoi{10.3847/1538-4357/ac3f38}

\bibitem[{{Keshet} {et~al.}(2009){Keshet}, {Katz}, {Spitkovsky}, \&
  {Waxman}}]{Keshet2009}
{Keshet}, U., {Katz}, B., {Spitkovsky}, A., \& {Waxman}, E. 2009, \apjl, 693,
  L127, \dodoi{10.1088/0004-637X/693/2/L127}

\bibitem[{{Khangulyan} {et~al.}(2021){Khangulyan}, {Aharonian}, {Romoli}, \&
  {Taylor}}]{Khangulyan2021}
{Khangulyan}, D., {Aharonian}, F., {Romoli}, C., \& {Taylor}, A. 2021, \apj,
  914, 76, \dodoi{10.3847/1538-4357/abfcbf}

\bibitem[{{Kirk} \& {Reville}(2010)}]{Kirk2010}
{Kirk}, J.~G., \& {Reville}, B. 2010, \apjl, 710, L16,
  \dodoi{10.1088/2041-8205/710/1/L16}

\bibitem[{{Kumar} {et~al.}(2012){Kumar}, {Hern{\'a}ndez}, {Bo{\v{s}}njak}, \&
  {Barniol Duran}}]{KumarP2012}
{Kumar}, P., {Hern{\'a}ndez}, R.~A., {Bo{\v{s}}njak}, {\v{Z}}., \& {Barniol
  Duran}, R. 2012, \mnras, 427, L40, \dodoi{10.1111/j.1745-3933.2012.01341.x}

\bibitem[{{Kumar} \& {Zhang}(2015)}]{KumarP2015}
{Kumar}, P., \& {Zhang}, B. 2015, \physrep, 561, 1,
  \dodoi{10.1016/j.physrep.2014.09.008}

\bibitem[{{Laskar} {et~al.}(2019){Laskar}, {van Eerten}, {Schady}, {Mundell},
  {Alexander}, {Barniol Duran}, {Berger}, {Bolmer}, {Chornock}, {Coppejans},
  {Fong}, {Gomboc}, {Jordana-Mitjans}, {Kobayashi}, {Margutti}, {Menten},
  {Sari}, {Yamazaki}, {Lipunov}, {Gorbovskoy}, {Kornilov}, {Tyurina},
  {Zimnukhov}, {Podesta}, {Levato}, {Buckley}, {Tlatov}, {Rebolo}, \&
  {Serra-Ricart}}]{Laskar2019}
{Laskar}, T., {van Eerten}, H., {Schady}, P., {et~al.} 2019, \apj, 884, 121,
  \dodoi{10.3847/1538-4357/ab40ce}

\bibitem[{{Laskar} {et~al.}(2023){Laskar}, {Alexander}, {Margutti},
  {Eftekhari}, {Chornock}, {Berger}, {Cendes}, {Duerr}, {Perley}, {Ravasio},
  {Yamazaki}, {Ayache}, {Barclay}, {Duran}, {Bhandari}, {Brethauer}, {Christy},
  {Coppejans}, {Duffell}, {Fong}, {Gomboc}, {Guidorzi}, {Kennea}, {Kobayashi},
  {Levan}, {Lobanov}, {Metzger}, {Ros}, {Schroeder}, \&
  {Williams}}]{Laskar2023}
{Laskar}, T., {Alexander}, K.~D., {Margutti}, R., {et~al.} 2023, \apjl, 946,
  L23, \dodoi{10.3847/2041-8213/acbfad}

\bibitem[{{Lemoine}(2015)}]{Lemoine2015}
{Lemoine}, M. 2015, Journal of Plasma Physics, 81, 455810101,
  \dodoi{10.1017/S0022377814000920}

\bibitem[{{Lemoine} {et~al.}(2019{\natexlab{a}}){Lemoine}, {Gremillet},
  {Pelletier}, \& {Vanthieghem}}]{Lemoine2019}
{Lemoine}, M., {Gremillet}, L., {Pelletier}, G., \& {Vanthieghem}, A.
  2019{\natexlab{a}}, \prl, 123, 035101, \dodoi{10.1103/PhysRevLett.123.035101}

\bibitem[{{Lemoine} \& {Pelletier}(2011)}]{Lemoine2011}
{Lemoine}, M., \& {Pelletier}, G. 2011, \mnras, 417, 1148,
  \dodoi{10.1111/j.1365-2966.2011.19331.x}

\bibitem[{{Lemoine} {et~al.}(2019{\natexlab{b}}){Lemoine}, {Pelletier},
  {Vanthieghem}, \& {Gremillet}}]{Lemoine2019c}
{Lemoine}, M., {Pelletier}, G., {Vanthieghem}, A., \& {Gremillet}, L.
  2019{\natexlab{b}}, \pre, 100, 033210, \dodoi{10.1103/PhysRevE.100.033210}

\bibitem[{{LHAASO Collaboration} {et~al.}(2023){LHAASO Collaboration}, {Cao},
  {Aharonian}, {An}, {Axikegu}, {Bai}, {Bai}, {Bao}, {Bastieri}, {Bi}, {Bi},
  {Cai}, {Cao}, {Cao}, {Cao}, {Chang}, {Chang}, {Chen}, {Chen}, {Chen}, {Chen},
  {Chen}, {Chen}, {Chen}, {Chen}, {Chen}, {Chen}, {Chen}, {Cheng}, {Cheng},
  {Cheng}, {Cui}, {Cui}, {Cui}, {Dai}, {Dai}, {Danzengluobu}, {Della Volpe},
  {Dong}, {Duan}, {Fan}, {Fan}, {Fang}, {Fang}, {Feng}, {Feng}, {Feng}, {Feng},
  {Feng}, {Gao}, {Gao}, {Gao}, {Gao}, {Gao}, {Gao}, {Ge}, {Geng}, {Gong},
  {Gou}, {Gu}, {Guo}, {Guo}, {Guo}, {Guo}, {Han}, {He}, {He}, {He}, {He}, {He},
  {Heller}, {Hor}, {Hou}, {Hou}, {Hou}, {Hu}, {Hu}, {Hu}, {Huang}, {Huang},
  {Huang}, {Huang}, {Huang}, {Ji}, {Jia}, {Jia}, {Jiang}, {Jiang}, {Jiang},
  {Jin}, {Kang}, {Ke}, {Kuleshov}, {Kurinov}, {Li}, {Li}, {Li}, {Li}, {Li},
  {Li}, {Li}, {Li}, {Li}, {Li}, {Li}, {Li}, {Li}, {Li}, {Li}, {Li}, {Li}, {Li},
  {Li}, {Liang}, {Liang}, {Lin}, {Liu}, {Liu}, {Liu}, {Liu}, {Liu}, {Liu},
  {Liu}, {Liu}, {Liu}, {Liu}, {Liu}, {Liu}, {Liu}, {Liu}, {Liu}, {Liu}, {Long},
  {Lu}, {Luo}, {Lv}, {Ma}, {Ma}, {Ma}, {Mao}, {Min}, {Mitthumsiri}, {Nan},
  {Ou}, {Pang}, {Pattarakijwanich}, {Pei}, {Qi}, {Qi}, {Qiao}, {Qin},
  {Ruffolo}, {Saiz}, {Shao}, {Shao}, {Shchegolev}, {Sheng}, {Song}, {Stenkin},
  {Stepanov}, {Su}, {Sun}, {Sun}, {Sun}, {Tam}, {Tang}, {Tian}, {Wang}, {Wang},
  {Wang}, {Wang}, {Wang}, {Wang}, {Wang}, {Wang}, {Wang}, {Wang}, {Wang},
  {Wang}, {Wang}, {Wang}, {Wang}, {Wang}, {Wang}, {Wang}, {Wang}, {Wei}, {Wei},
  {Wei}, {Wen}, {Wu}, {Wu}, {Wu}, {Wu}, {Wu}, {Xi}, {Xia}, {Xia}, {Xiang},
  {Xiao}, {Xiao}, {Xin}, {Xin}, {Xing}, {Xiong}, {Xu}, {Xu}, {Xu}, {Xue},
  {Yan}, {Yan}, {Yan}, {Yang}, {Yang}, {Yang}, {Yang}, {Yang}, {Yang}, {Yang},
  {Yang}, {Yang}, {Yao}, {Ye}, {Yin}, {Yin}, {You}, {You}, {Yu}, {Yuan}, {Yue},
  {Zeng}, {Zeng}, {Zeng}, {Zeng}, {Zhang}, {Zhang}, {Zhang}, {Zhang}, {Zhang},
  {Zhang}, {Zhang}, {Zhang}, {Zhang}, {Zhang}, {Zhang}, {Zhang}, {Zhang},
  {Zhang}, {Zhang}, {Zhang}, {Zhang}, {Zhang}, {Zhang}, {Zhao}, {Zhao}, {Zhao},
  {Zhao}, {Zhao}, {Zheng}, {Zhou}, {Zhou}, {Zhou}, {Zhou}, {Zhou}, {Zhou},
  {Zhu}, {Zhu}, {Zhu}, {Zhu}, \& {Zuo}}]{LHAASO2023}
{LHAASO Collaboration}, {Cao}, Z., {Aharonian}, F., {et~al.} 2023, Science,
  380, 1390, \dodoi{10.1126/science.adg9328}

\bibitem[{{MAGIC Collaboration} {et~al.}(2019{\natexlab{a}}){MAGIC
  Collaboration}, {Acciari}, {Ansoldi}, {Antonelli}, {Engels}, {Baack},
  {Babi{\'c}}, {Banerjee}, {Barres de Almeida}, {Barrio}, {Becerra
  Gonz{\'a}lez}, {Bednarek}, {Bellizzi}, {Bernardini}, {Berti}, {Besenrieder},
  {Bhattacharyya}, {Bigongiari}, {Biland}, {Blanch}, {Bonnoli},
  {Bo{\v{s}}njak}, {Busetto}, {Carosi}, {Ceribella}, {Chai}, {Chilingaryan},
  {Cikota}, {Colak}, {Colin}, {Colombo}, {Contreras}, {Cortina}, {Covino},
  {D'Elia}, {da Vela}, {Dazzi}, {de Angelis}, {de Lotto}, {Delfino}, {Delgado},
  {Depaoli}, {di Pierro}, {di Venere}, {Do Souto Espi{\~n}eira}, {Dominis
  Prester}, {Donini}, {Dorner}, {Doro}, {Elsaesser}, {Fallah Ramazani},
  {Fattorini}, {Ferrara}, {Fidalgo}, {Foffano}, {Fonseca}, {Font}, {Fruck},
  {Fukami}, {Garc{\'\i}a L{\'o}pez}, {Garczarczyk}, {Gasparyan}, {Gaug},
  {Giglietto}, {Giordano}, {Godinovi{\'c}}, {Green}, {Guberman}, {Hadasch},
  {Hahn}, {Herrera}, {Hoang}, {Hrupec}, {H{\"u}tten}, {Inada}, {Inoue},
  {Ishio}, {Iwamura}, {Jouvin}, {Kerszberg}, {Kubo}, {Kushida}, {Lamastra},
  {Lelas}, {Leone}, {Lindfors}, {Lombardi}, {Longo}, {L{\'o}pez},
  {L{\'o}pez-Coto}, {L{\'o}pez-Oramas}, {Loporchio}, {Machado de Oliveira
  Fraga}, {Maggio}, {Majumdar}, {Makariev}, {Mallamaci}, {Maneva}, {Manganaro},
  {Mannheim}, {Maraschi}, {Mariotti}, {Mart{\'\i}nez}, {Mazin},
  {Mi{\'c}anovi{\'c}}, {Miceli}, {Minev}, {Miranda}, {Mirzoyan}, {Molina},
  {Moralejo}, {Morcuende}, {Moreno}, {Moretti}, {Munar-Adrover}, {Neustroev},
  {Nigro}, {Nilsson}, {Ninci}, {Nishijima}, {Noda}, {Nogu{\'e}s}, {Nozaki},
  {Paiano}, {Palatiello}, {Paneque}, {Paoletti}, {Paredes}, {Pe{\~n}il},
  {Peresano}, {Persic}, {Moroni}, {Prandini}, {Puljak}, {Rhode}, {Rib{\'o}},
  {Rico}, {Righi}, {Rugliancich}, {Saha}, {Sahakyan}, {Saito}, {Sakurai},
  {Satalecka}, {Schmidt}, {Schweizer}, {Sitarek}, {{\v{S}}nidari{\'c}},
  {Sobczynska}, {Somero}, {Stamerra}, {Strom}, {Strzys}, {Suda}, {Suri{\'c}},
  {Takahashi}, {Tavecchio}, {Temnikov}, {Terzi{\'c}}, {Teshima},
  {Torres-Alb{\`a}}, {Tosti}, {Vagelli}, {van Scherpenberg}, {Vanzo}, {Vazquez
  Acosta}, {Vigorito}, {Vitale}, {Vovk}, {Will}, {Zari{\'c}}, {Nava}, {Veres},
  {Bhat}, {Briggs}, {Cleveland}, {Hamburg}, {Hui}, {Mailyan}, {Preece},
  {Roberts}, {von Kienlin}, {Wilson-Hodge}, {Kocevski}, {Arimoto}, {Tak},
  {Asano}, {Axelsson}, {Barbiellini}, {Bissaldi}, {Dirirsa}, {Gill}, {Granot},
  {McEnery}, {Omodei}, {Razzaque}, {Piron}, {Racusin}, {Thompson}, {Campana},
  {Bernardini}, {Kuin}, {Siegel}, {Cenko}, {O'Brien}, {Capalbi}, {Da{\i}}, {de
  Pasquale}, {Gropp}, {Klingler}, {Osborne}, {Perri}, {Starling},
  {Tagliaferri}, {Tohuvavohu}, {Ursi}, {Tavani}, {Cardillo}, {Casentini},
  {Piano}, {Evangelista}, {Verrecchia}, {Pittori}, {Lucarelli}, {Bulgarelli},
  {Parmiggiani}, {Anderson}, {Anderson}, {Bernardi}, {Bolmer},
  {Caballero-Garc{\'\i}a}, {Carrasco}, {Castell{\'o}n}, {Castro Segura},
  {Castro-Tirado}, {Cherukuri}, {Cockeram}, {D'Avanzo}, {di Dato}, {Diretse},
  {Fender}, {Fern{\'a}ndez-Garc{\'\i}a}, {Fynbo}, {Fruchter}, {Greiner},
  {Gromadzki}, {Heintz}, {Heywood}, {van der Horst}, {Hu}, {Inserra}, {Izzo},
  {Jaiswal}, {Jakobsson}, {Japelj}, {Kankare}, {Kann}, {Kouveliotou}, {Klose},
  {Levan}, {Li}, {Lotti}, {Maguire}, {Malesani}, {Manulis}, {Marongiu},
  {Martin}, {Melandri}, {Micha{\l}owski}, {Miller-Jones}, {Misra}, {Moin},
  {Mooley}, {Nasri}, {Nicholl}, {Noschese}, {Novara}, {Pandey}, {Peretti},
  {P{\'e}rez Del Pulgar}, {P{\'e}rez-Torres}, {Perley}, {Piro}, {Ragosta},
  {Resmi}, {Ricci}, {Rossi}, {S{\'a}nchez-Ram{\'\i}rez}, {Selsing}, {Schulze},
  {Smartt}, {Smith}, {Sokolov}, {Stevens}, {Tanvir}, {Th{\"o}ne}, {Tiengo},
  {Tremou}, {Troja}, {de Ugarte Postigo}, {Valeev}, {Vergani}, {Wieringa},
  {Woudt}, {Xu}, {Yaron}, \& {Young}}]{MAGIC2019}
{MAGIC Collaboration}, {Acciari}, V.~A., {Ansoldi}, S., {et~al.}
  2019{\natexlab{a}}, \nat, 575, 459, \dodoi{10.1038/s41586-019-1754-6}

\bibitem[{{MAGIC Collaboration} {et~al.}(2019{\natexlab{b}}){MAGIC
  Collaboration}, {Acciari}, {Ansoldi}, {Antonelli}, {Arbet Engels}, {Baack},
  {Babi{\'c}}, {Banerjee}, {Barres de Almeida}, {Barrio}, {Becerra
  Gonz{\'a}lez}, {Bednarek}, {Bellizzi}, {Bernardini}, {Berti}, {Besenrieder},
  {Bhattacharyya}, {Bigongiari}, {Biland}, {Blanch}, {Bonnoli},
  {Bo{\v{s}}njak}, {Busetto}, {Carosi}, {Carosi}, {Ceribella}, {Chai},
  {Chilingaryan}, {Cikota}, {Colak}, {Colin}, {Colombo}, {Contreras},
  {Cortina}, {Covino}, {D'Amico}, {D'Elia}, {da Vela}, {Dazzi}, {de Angelis},
  {de Lotto}, {Delfino}, {Delgado}, {Depaoli}, {di Pierro}, {di Venere}, {Do
  Souto Espi{\~n}eira}, {Dominis Prester}, {Donini}, {Dorner}, {Doro},
  {Elsaesser}, {Fallah Ramazani}, {Fattorini}, {Fern{\'a}ndez-Barral},
  {Ferrara}, {Fidalgo}, {Foffano}, {Fonseca}, {Font}, {Fruck}, {Fukami},
  {Gallozzi}, {Garc{\'\i}a L{\'o}pez}, {Garczarczyk}, {Gasparyan}, {Gaug},
  {Giglietto}, {Giordano}, {Godinovi{\'c}}, {Green}, {Guberman}, {Hadasch},
  {Hahn}, {Herrera}, {Hoang}, {Hrupec}, {H{\"u}tten}, {Inada}, {Inoue},
  {Ishio}, {Iwamura}, {Jouvin}, {Kerszberg}, {Kubo}, {Kushida}, {Lamastra},
  {Lelas}, {Leone}, {Lindfors}, {Lombardi}, {Longo}, {L{\'o}pez},
  {L{\'o}pez-Coto}, {L{\'o}pez-Oramas}, {Loporchio}, {Machado de Oliveira
  Fraga}, {Maggio}, {Majumdar}, {Makariev}, {Mallamaci}, {Maneva}, {Manganaro},
  {Mannheim}, {Maraschi}, {Mariotti}, {Mart{\'\i}nez}, {Masuda}, {Mazin},
  {Mi{\'c}anovi{\'c}}, {Miceli}, {Minev}, {Miranda}, {Mirzoyan}, {Molina},
  {Moralejo}, {Morcuende}, {Moreno}, {Moretti}, {Munar-Adrover}, {Neustroev},
  {Nigro}, {Nilsson}, {Ninci}, {Nishijima}, {Noda}, {Nogu{\'e}s}, {N{\"o}the},
  {Nozaki}, {Paiano}, {Palacio}, {Palatiello}, {Paneque}, {Paoletti},
  {Paredes}, {Pe{\~n}il}, {Peresano}, {Persic}, {Prada Moroni}, {Prandini},
  {Puljak}, {Rhode}, {Rib{\'o}}, {Rico}, {Righi}, {Rugliancich}, {Saha},
  {Sahakyan}, {Saito}, {Sakurai}, {Satalecka}, {Schmidt}, {Schweizer},
  {Sitarek}, {{\v{S}}nidari{\'c}}, {Sobczynska}, {Somero}, {Stamerra}, {Strom},
  {Strzys}, {Suda}, {Suri{\'c}}, {Takahashi}, {Tavecchio}, {Temnikov},
  {Terzi{\'c}}, {Teshima}, {Torres-Alb{\`a}}, {Tosti}, {Tsujimoto}, {Vagelli},
  {van Scherpenberg}, {Vanzo}, {Vazquez Acosta}, {Vigorito}, {Vitale}, {Vovk},
  {Will}, {Zari{\'c}}, \& {Nava}}]{MAGIC2019b}
---. 2019{\natexlab{b}}, \nat, 575, 455, \dodoi{10.1038/s41586-019-1750-x}

\bibitem[{{Martins} {et~al.}(2009){Martins}, {Fonseca}, {Silva}, \&
  {Mori}}]{Martins2009}
{Martins}, S.~F., {Fonseca}, R.~A., {Silva}, L.~O., \& {Mori}, W.~B. 2009,
  \apjl, 695, L189, \dodoi{10.1088/0004-637X/695/2/L189}

\bibitem[{{Medvedev} \& {Loeb}(1999)}]{Medvedev1999}
{Medvedev}, M.~V., \& {Loeb}, A. 1999, \apj, 526, 697, \dodoi{10.1086/308038}

\bibitem[{{M{\'e}sz{\'a}ros}(2002)}]{Meszaros2002}
{M{\'e}sz{\'a}ros}, P. 2002, \araa, 40, 137,
  \dodoi{10.1146/annurev.astro.40.060401.093821}

\bibitem[{{M{\'e}sz{\'a}ros}(2006)}]{Meszaros2006}
---. 2006, Reports on Progress in Physics, 69, 2259,
  \dodoi{10.1088/0034-4885/69/8/R01}

\bibitem[{{Nishikawa} {et~al.}(2009){Nishikawa}, {Niemiec}, {Hardee},
  {Medvedev}, {Sol}, {Mizuno}, {Zhang}, {Pohl}, {Oka}, \&
  {Hartmann}}]{Nishikawa2009}
{Nishikawa}, K.~I., {Niemiec}, J., {Hardee}, P.~E., {et~al.} 2009, \apjl, 698,
  L10, \dodoi{10.1088/0004-637X/698/1/L10}

\bibitem[{{Panaitescu} \& {Kumar}(2002)}]{Panaitescu2002}
{Panaitescu}, A., \& {Kumar}, P. 2002, \apj, 571, 779, \dodoi{10.1086/340094}

\bibitem[{{Parsons} {et~al.}(2023){Parsons}, {Spitkovsky}, \&
  {Vanthieghem}}]{Parsons2023}
{Parsons}, J., {Spitkovsky}, A., \& {Vanthieghem}, A. 2023, arXiv e-prints,
  arXiv:2310.12950, \dodoi{10.48550/arXiv.2310.12950}

\bibitem[{{Perley} {et~al.}(2014){Perley}, {Cenko}, {Corsi}, {Tanvir}, {Levan},
  {Kann}, {Sonbas}, {Wiersema}, {Zheng}, {Zhao}, {Bai}, {Bremer},
  {Castro-Tirado}, {Chang}, {Clubb}, {Frail}, {Fruchter},
  {G{\"o}{\u{g}}{\"u}{\c{s}}}, {Greiner}, {G{\"u}ver}, {Horesh}, {Filippenko},
  {Klose}, {Mao}, {Morgan}, {Pozanenko}, {Schmidl}, {Stecklum}, {Tanga},
  {Volnova}, {Volvach}, {Wang}, {Winters}, \& {Xin}}]{Perley2014}
{Perley}, D.~A., {Cenko}, S.~B., {Corsi}, A., {et~al.} 2014, \apj, 781, 37,
  \dodoi{10.1088/0004-637X/781/1/37}

\bibitem[{{Peterson} {et~al.}(2021){Peterson}, {Glenzer}, \&
  {Fiuza}}]{Peterson2021}
{Peterson}, J.~R., {Glenzer}, S., \& {Fiuza}, F. 2021, \prl, 126, 215101,
  \dodoi{10.1103/PhysRevLett.126.215101}

\bibitem[{{Peterson} {et~al.}(2022){Peterson}, {Glenzer}, \&
  {Fiuza}}]{Peterson2021b}
---. 2022, \apjl, 924, L12, \dodoi{10.3847/2041-8213/ac44a2}

\bibitem[{{Piran}(2004)}]{Piran2004}
{Piran}, T. 2004, Reviews of Modern Physics, 76, 1143,
  \dodoi{10.1103/RevModPhys.76.1143}

\bibitem[{{Plotnikov} {et~al.}(2018){Plotnikov}, {Grassi}, \&
  {Grech}}]{Plotnikov2018}
{Plotnikov}, I., {Grassi}, A., \& {Grech}, M. 2018, \mnras, 477, 5238,
  \dodoi{10.1093/mnras/sty979}

\bibitem[{{Plotnikov} {et~al.}(2011){Plotnikov}, {Pelletier}, \&
  {Lemoine}}]{Plotnikov2011}
{Plotnikov}, I., {Pelletier}, G., \& {Lemoine}, M. 2011, \aap, 532, A68,
  \dodoi{10.1051/0004-6361/201117182}

\bibitem[{{Ressler} \& {Laskar}(2017)}]{Ressler2017}
{Ressler}, S.~M., \& {Laskar}, T. 2017, \apj, 845, 150,
  \dodoi{10.3847/1538-4357/aa8268}

\bibitem[{{Reville} \& {Bell}(2014)}]{Reville2014}
{Reville}, B., \& {Bell}, A.~R. 2014, \mnras, 439, 2050,
  \dodoi{10.1093/mnras/stu088}

\bibitem[{{Sari} {et~al.}(1998){Sari}, {Piran}, \& {Narayan}}]{Sari1998}
{Sari}, R., {Piran}, T., \& {Narayan}, R. 1998, \apjl, 497, L17,
  \dodoi{10.1086/311269}

\bibitem[{{Silva} {et~al.}(2003){Silva}, {Fonseca}, {Tonge}, {Dawson}, {Mori},
  \& {Medvedev}}]{Silva2003}
{Silva}, L.~O., {Fonseca}, R.~A., {Tonge}, J.~W., {et~al.} 2003, \apjl, 596,
  L121, \dodoi{10.1086/379156}

\bibitem[{{Sironi} {et~al.}(2013){Sironi}, {Spitkovsky}, \&
  {Arons}}]{Sironi2013}
{Sironi}, L., {Spitkovsky}, A., \& {Arons}, J. 2013, \apj, 771, 54,
  \dodoi{10.1088/0004-637X/771/1/54}

\bibitem[{{Spitkovsky}(2008{\natexlab{a}})}]{Spitkovsky2008}
{Spitkovsky}, A. 2008{\natexlab{a}}, \apjl, 682, L5, \dodoi{10.1086/590248}

\bibitem[{{Spitkovsky}(2008{\natexlab{b}})}]{Spitkovsky2008b}
---. 2008{\natexlab{b}}, \apjl, 673, L39, \dodoi{10.1086/527374}

\bibitem[{{Takamoto} {et~al.}(2018){Takamoto}, {Matsumoto}, \&
  {Kato}}]{Takamoto2018}
{Takamoto}, M., {Matsumoto}, Y., \& {Kato}, T.~N. 2018, \apjl, 860, L1,
  \dodoi{10.3847/2041-8213/aac6d6}

\bibitem[{{Warren} {et~al.}(2022){Warren}, {Dainotti}, {Barkov}, {Ahlgren},
  {Ito}, \& {Nagataki}}]{Warren2022}
{Warren}, D.~C., {Dainotti}, M., {Barkov}, M.~V., {et~al.} 2022, \apj, 924, 40,
  \dodoi{10.3847/1538-4357/ac2f43}

\bibitem[{{Weibel}(1959)}]{Weibel1959}
{Weibel}, E.~S. 1959, \prl, 2, 83, \dodoi{10.1103/PhysRevLett.2.83}

\bibitem[{{Wijers} \& {Galama}(1999)}]{Wijers1999}
{Wijers}, R.~A.~M.~J., \& {Galama}, T.~J. 1999, \apj, 523, 177,
  \dodoi{10.1086/307705}

\bibitem[{{Williams} {et~al.}(2023){Williams}, {Kennea}, {Dichiara},
  {Kobayashi}, {Iwakiri}, {Beardmore}, {Evans}, {Heinz}, {Lien}, {Oates},
  {Negoro}, {Cenko}, {Buisson}, {Hartmann}, {Jaisawal}, {Kuin}, {Lesage},
  {Page}, {Parsotan}, {Pasham}, {Sbarufatti}, {Siegel}, {Sugita}, {Younes},
  {Ambrosi}, {Arzoumanian}, {Bernardini}, {Campana}, {Capalbi}, {Caputo},
  {D'A{\`\i}}, {D'Avanzo}, {D'Elia}, {De Pasquale}, {Eyles-Ferris}, {Ferrara},
  {Gendreau}, {Gropp}, {Kawai}, {Klingler}, {Laha}, {Melandri}, {Mihara},
  {Moss}, {O'Brien}, {Osborne}, {Palmer}, {Perri}, {Serino}, {Sonbas},
  {Stamatikos}, {Starling}, {Tagliaferri}, {Tohuvavohu}, {Zane}, \&
  {Ziaeepour}}]{Williams2023}
{Williams}, M.~A., {Kennea}, J.~A., {Dichiara}, S., {et~al.} 2023, \apjl, 946,
  L24, \dodoi{10.3847/2041-8213/acbcd1}

\end{thebibliography}

\end{document}